\documentclass[a4paper,twoside,12pt]{article}

\usepackage{xspace}
\usepackage{amsmath}
\usepackage{amscd}
\usepackage{amssymb}
\usepackage{epic}
\usepackage{eepic}
\usepackage{pifont}
\usepackage{rotating}
\usepackage{color}

\renewcommand{\thefootnote}{}

\title{On Decidability of Expressive Description Logics\\
with Composition of Roles in Number Restrictions
}
\author{\bf Fabio Grandi\\
IEIIT.BO-CNR and DEIS,\\
Alma Mater Studiorum -- Universit\`a di Bologna,\\
Viale Risorgimento 2, I-40136 Bologna, Italy\\
Email: {\tt fgrandi@deis.unibo.it}}
\date{}

\newcommand{\QED}{\hfill$\Box$}

\newcommand{\Z}{\mathbb{Z}}

\newcommand{\NI}{\textsf{NI}}
\newcommand{\NC}{\textsf{NC}}
\newcommand{\NR}{\textsf{NR}}

\newcommand{\choos}{\Pisymbol{psy}{227}}

\newcommand{\set}[1]{\{ #1 \}}          

\newcommand{\Cdue}{\ALlang{C}{^2}}

\newcommand{\ALlang}[2]{\ensuremath{\mathcal{#1}#2}\xspace}

\newcommand{\ALC}{\ALlang{ALC}{}}
\newcommand{\ALCN}{\ALlang{ALCN}{}}
\newcommand{\ALCQ}{\ALlang{ALCQ}{}}

\newcommand{\ALCNoI}{\ALlang{ALCN}{(\comp)\negmedspace}\ALlang{I}{}}
\newcommand{\ALCNX}[1]{\ALlang{ALCN}{(#1)}}

\newcommand{\ALCNNX}[1]{\ALlang{ALC\bar N}{(#1)}}
\newcommand{\ALCQX}[1]{\ALlang{ALCQ}{(#1)}}

\newcommand{\ALCpN}{\ALlang{ALC}$_+$\ALlang{N}{}}

\newcommand{\ALCregN}{\ALlang{ALC}{_{\text{reg}}\ALlang{N}}}

\newcommand{\PSPACE}{\textsc{PSpace}\xspace}
\newcommand{\NPSPACE}{\textsc{NPSpace}\xspace}

\newcommand{\NEXPTIME}{\textsc{NExpTime}\xspace}
\newcommand{\DNEXPTIME}{\textsc{2-NExpTime}\xspace}

\newcommand{\I}{\mathcal{I}}           
\newcommand{\dom}[1][\I]{\Delta^{#1}}  
\newcommand{\Int}[2][\I]{#2^{#1}}      

\newcommand{\Implies}{\sqsubseteq}                   

\newcommand{\Bot}{\ensuremath{\bot}}
\newcommand{\Top}{\ensuremath{\top}}

\renewcommand{\And}{\ensuremath{\sqcap}}
\newcommand{\Or}{\ensuremath{\sqcup}}
\newcommand{\Not}{\ensuremath{\neg}}

\newcommand{\some}[2]{\ensuremath{\exists} #1 . #2}
\newcommand{\all}[2]{\ensuremath{\forall} #1 . #2}
\newcommand{\atleast}[2]{\ensuremath{\exists^{\geq #1}} #2}
\newcommand{\atmost}[2]{\ensuremath{\exists^{\leq #1}} #2}

\newcommand{\atleastq}[3]{\ensuremath{\exists^{\geq #1}} #2 . #3}
\newcommand{\atmostq}[3]{\ensuremath{\exists^{\leq #1}} #2 . #3}

\newcommand{\inv}[1]{#1^{-}}              
\newcommand{\comp}{\circ}                 

\renewcommand{\ding}[1]{\text{\Pisymbol{pzd}{#1}}}

\newtheorem{theorem}{Theorem}
\newtheorem{corollary}{Corollary}
\newtheorem{Definition}{Definition}

\newtheorem{lemma}{Lemma}
\newtheorem{new_example}{Example}

\newenvironment{definition}{\begin{Definition} \rm}
{\end{Definition}}

\def\|{\ensuremath{\ \mid}}
\def\per{\ensuremath{\mathbf .\ }\xspace}

\newcommand{\xput}[2]{\put(#1){\makebox(0,0){$#2$}}}

\newcommand{\biggOr}[1]{\underset{#1}{\text{\begin{huge}$\sqcup$\end{huge}}}}
\newcommand{\biggAnd}[1]{\underset{#1}{\text{\begin{huge}$\sqcap$\end{huge}}}}

\newcommand{\bigOr}[1]{\underset{#1}{\text{\begin{Large}$\sqcup$\end{Large}}}}
\newcommand{\bigAnd}[1]{\underset{#1}{\text{\begin{Large}$\sqcap$\end{Large}}}}

\newcommand{\level}{\textsl{level}}
\newcommand{\depth}{\textsl{depth}}

\newcommand{\A}{\ensuremath{{\cal A}}}

\newcommand{\atbothq}[3]{\exists^{\Join#1}#2.#3}

\begin{document}

\maketitle

\begin{abstract}
Description Logics are knowledge representation formalisms
which have been used in a wide range of application domains.
Owing to their appealing expressiveness, we consider in this paper
extensions of the well-known concept language \ALC allowing for
\emph{number restrictions} on complex role expressions.
These have been first introduced by Baader and Sattler as
\ALCNX{M} languages, with the adoption
of role constructors $M\subseteq \set{\comp,\inv{},\Or,\And}$.
In particular, they showed in 1999 that, although \ALCNX{\comp}
is decidable, the addition of other operators may easily lead
to undecidability: in fact, \ALCNX{\comp,\And} and \ALCNX{\comp,\inv{},\Or}
were proved undecidable.

In this work, we further investigate the computational properties
of the \ALCN\ family, aiming at narrowing the decidability gap left open
by Baader and Sattler's results.
In particular, we will show that
\ALCNX{\comp} extended with inverse roles both in number and in value restrictions
becomes undecidable, whereas it can be safely extended
with qualified number restrictions without losing decidability.
\end{abstract}

%

\noindent \textbf{Keywords:} Description Logic, Decidability, Domino Problem, Tableau Algorithm

\section{Introduction}

Description Logics (DLs) are a family of first-order formalisms that
have been found useful for domain knowledge representation in several
application fields \cite{DLbook}, from database design ---including conceptual, object-oriented,
temporal, multimedia and semistructured data modeling--- to software engineering
and ontology management (e.g. \cite{AF99,BJ92,GHB96,JLC-DL-99,KR-98,DBIS98,franconi-dood00,HS01}).
Different DLs provide for \emph{constructors} which can be used to
combine atomic \emph{concepts} (unary predicates) and \emph{roles} (binary predicates)
to build \emph{complex} concepts and roles. The available constructors
characterize the description language as to \emph{expressiveness} and
\emph{computational behaviour} (decidability and complexity) of the basic reasoning tasks
like concept satisfiability and subsumption.

\footnotetext{This is the extended version of a paper which appeared in
\emph{Proceedings of LPAR~2002 -- 9th Intl' Conf. on Logic for Programming,
Artificial Intelligence and Reasoning} (Tbilisi, Georgia, October 2002).}
\renewcommand{\thefootnote}{\arabic{footnote}}
\clearpage

Well-known Description Logics are \ALC \cite{ALC}, which allows for Boolean
propositional constructors on concepts and (universal and existential) value restrictions
on atomic roles, and its extension \ALCN \cite{ALCN1,ALCN2} introducing (non-qualified)
\emph{number restrictions} on atomic roles. Basic inference problems for both
these DLs are \PSPACE-complete \cite{ALCN1,ALCN2}.
However, in order to better fulfil requirements of real-world application domains,
more expressive extensions of the basic concept languages have been investigated.
One direction along which useful extensions
have been sought is the introduction of \emph{complex} roles under number restrictions.
In fact, considering role composition $(\comp)$, inversion $(^-)$, union $(\Or)$ and intersection $(\And)$,
expressive extensions of \ALCN\ can be defined as \ALCNX{M} with the adoption
of role constructors $M\subseteq \set{\comp,\inv{},\Or,\And}$ \cite{BS99}.
By allowing (different kinds of) complex roles also in
value restrictions, 
different families of Logics can also be defined: for example \ALCpN (or \ALCregN) allows the
transitive closure of atomic roles (or regular roles, resp.) under value restrictions \cite{BS99,dl-expr-reas:book-99}.
Also Logics \ALCNNX{M}, allowing for the same types of role constructors
either in value \emph{and} in number restrictions, can be considered \cite{myDL01}.
Further extensions involve the introduction of \emph{qualified} number restrictions \cite{HB91}
on complex roles, giving rise to \ALCQX{M} Logics.
Since qualified number restrictions also allow to express value restrictions,
we have the inclusions $\ALCNX{M}\subseteq \ALCNNX{M}\subseteq \ALCQX{M}$ as far as
expressiveness (and complexity) are concerned.
Therefore, for instance, undecidability of \ALCNX{M} directly extends to \ALCNNX{M} and \ALCQX{M},
whereas decidability of \ALCQX{M} implies decidability of \ALCNX{M} and \ALCNNX{M}.
\begin{figure}[t]
{\small
\begin{center}
    $\begin{array}%
    {r@{\hspace*{+.5ex}}c@{\hspace*{+.5ex}}l@{\hspace*{1em}}rcl}
      C,D & \rightarrow\
        & A \mid &  \Int A\subseteq\Int\Delta && \text{atomic concept}\\
      & & \Top \mid & \Int\Top &=& \Int\Delta \\
      & & \Bot \mid & \Int\Bot &=& \emptyset \\
      & & \Not C  \mid & \Int{(\Not C)} &=& \Int\Delta \setminus \Int C \\
      & & C\And D \mid & \Int{(C\And D)} &=& \Int C \cap \Int D \\
      & & C\Or D \mid & \Int{(C\Or D)} &=& \Int C \cup \Int D\\
      & & \all{R}{C} \mid & \Int{(\all{R}{C})} &=&
          \set{i\in\Int\Delta\mid\forall j\per\Int R(i,j)\Rightarrow\Int C(j)}\\
      & & \some{R}{C} \mid & \Int{(\some{R}{C})} &=&
          \set{i\in\Int\Delta\mid\exists j\per\Int R(i,j)\wedge\Int C(j)} \\
      & & \atleast{n}{R} \mid & \Int{(\atleast{n}{R})} &=&
          \set{i\in\Int\Delta\mid\sharp
          \set{j\in\Int\Delta\mid \Int R(i,j)} \geq n} \\
      & & \atmost{n}{R} \mid & \Int{(\atmost{n}{R})} &=&
          \set{i\in\Int\Delta\mid\sharp
          \set{j\in\Int\Delta\mid \Int R(i,j)} \leq n} \\
      & * & \atleastq{n}{R}{C} \mid & \Int{(\atleastq{n}{R}{C})} &=&
          \set{i\in\Int\Delta\mid\sharp
          \set{j\in\Int\Delta\mid \Int R(i,j)\wedge \Int C(j)} \geq n} \\
      & * & \atmostq{n}{R}{C} \mid & \Int{(\atmostq{n}{R}{C})} &=&
          \set{i\in\Int\Delta\mid\sharp
          \set{j\in\Int\Delta\mid \Int R(i,j)\wedge \Int C(j)} \leq n} \\
      R, S & \rightarrow\
        & P \mid & \Int P\subseteq\Int\Delta \times\Int\Delta && \text{atomic role} \\
      & * & \inv{R} \mid & \Int{(\inv{R})}  &=&  \set{ (i,j)\in\Int\Delta\times\Int\Delta\mid
           \Int R(j,i) } \\
      & * & R \comp S \mid & \Int{(R\comp S)} &=&  \set{(i,j)\in\Int\Delta\times\Int\Delta\mid
          \exists k\per \Int R(i,k) \wedge \Int S(k,j) } \\
    \end{array}$
\end{center}\vspace{-1em}
} \caption{\label{ALCN} Syntax and model-theoretic semantics of \ALCN
and its extensions (marked with $\ast$) considered in this paper.
}
\end{figure}


Our investigation is aimed at improving the (un)decidability results presented
by Baader and Sattler in \cite{BS99} for \ALCN\ extensions including composition of roles ($\comp$).
In particular, they proved that concept satisfiability in \ALCNX{\comp,\And} and \ALCNX{\comp,^-,\Or}
is undecidable via reduction of a domino problem,
and provided a sound and complete
Tableau algorithm for deciding satisfiability of \ALCNX{\comp}-concepts.
Furthermore, we recently proved that concept satisfiability is decidable in \ALCNX{\comp,\Or},
by providing a Tableau algorithm for the purpose \cite{myDL03}.
Moreover, it can easily be proved (using the Role Normal Form introduced in \cite{myDL03}) that
\ALCNNX{\comp,\Or} is simply a syntactic variant of \ALCNX{\comp,\Or}.
Baader and Sattler also observed in \cite{BS99} that \ALCNX{^-,\Or,\And} 
is decidable since \ALCNX{^-,\Or,\And}-concepts
can easily be translated into a formula in \Cdue\ \cite{C2}, that is
the two-variable FOL fragment with counting quantifiers, which has proved to be
decidable \cite{decC2}. In fact, satisfiability of \Cdue formulae can be decided
in \NEXPTIME \cite{complC2} if unary coding of numbers is used (which is a common assumption in the field
of DLs; if binary coding is adopted we have a \DNEXPTIME upper bound).
We can further observe that a similar translation is still possible
when \emph{qualified} number restrictions   
are considered and, thus, also  \ALCQX{^-,\Or,\And} and \ALCNNX{^-,\Or,\And}
are decidable.


In this paper, we consider extensions of \ALCNX{\comp}
with role inversion ($\cal I$) or qualified number restrictions ($\cal Q$),
whose decidability status, to the best of our knowledge, is still unknown.
In particular, we will show in Sec.~\ref{undec1} (via reduction of a domino problem)
undecidability of \ALCNX{\comp} extended with inverse roles
both in value and in number restrictions (which we can call \ALCNoI, but
we also show in Sec.~\ref{undec1} that it is a syntactic variant of \ALCNNX{\comp,^-})
is undecidable.
This result implies undecidability of \ALCQX{\comp,^-},
whereas decidability of ``pure'' \ALCNX{\comp,^-} remains an open question.
On the other hand, we will show how the decidability results of \cite{BS99}
lift up to \ALCQX{\comp}. In particular, we will show in Sec.~\ref{dec2}
that \ALCQX{\comp}-concept satisfiability is decidable and provide
an effective decision procedure in the form of a tableau-based algorithm,
which extends the \ALCNX{\comp} Tableau proposed by Baader and Sattler \cite{BS99}.
In a similar way as done in \cite{BS99}, we will also show that the
decision algorithm can be extended to cope with qualified number restrictions
on union and/or intersections of role chains of the same length.
Conclusions will eventually be found in Section~4.

\subsection*{Preliminaries on Description Logics}

Description Logics expressiveness is based on the definition
of complex concepts and roles, which can be built with the help
of available constructors, starting from a set of (atomic) concept names \NC\
and a set of (atomic) role names \NR. A DL system, enabling concept descriptions
to be interrelated, allows the derivation of implicit knowledge from
explicitly represented knowledge by means of inference services.
For a full account of Description Logics, the reader is referred, for example,
to \cite{DLbook}.

In the DL \ALC \cite{ALC}, concept descriptions are formed using the constructors
negation, conjunction and disjunction, value (and existential) restrictions.
The DL \ALCN \cite{ALCN1,ALCN2} additionally allows for unqualified (at-least and at-most)
number restrictions on atomic roles.
The syntax rules at the left hand side of Fig.~\ref{ALCN} inductively define
valid concept and role expressions for \ALCN and its extensions
considered in this paper.
As far as semantics is concerned, concepts are interpreted as sets of individuals
and roles as sets of pairs of individuals. Formally, an \emph{interpretation} is
a pair $\I=(\Int\Delta, \Int\cdot )$, where $\Int\Delta$ is a non-empty set of individuals
(the \emph{domain} of $\I$) and $\I$ is a function (the \emph{interpretation function})
which maps each concept to a subset of $\Int\Delta$ and each role to a subset of $\Int\Delta\times\Int\Delta$,
such that the equations at the right hand side of Fig.~\ref{ALCN} are satisfied.
One of the most important inference services of DL systems used in knowledge-representation
and conceptual modeling applications is computing the subsumption hierarchy
of a given finite set of concept descriptions.
\begin{definition}
The concept description $C$ is \emph{satisfiable} iff there exist an interpretation $\I$ such that
$\Int C \neq\emptyset$; in this case, we say that $\I$ is a model for $C$.
The concept description $D$ \emph{subsumes} the concept description $C$
(written $C\Implies D$) iff $\Int I\subseteq\Int D$ for all interpretations $\I$; concept
descriptions $C$ and $D$ are \emph{equivalent} iff $C\Implies D$ and $D\Implies C$.
\end{definition}
Since \ALC\ is propositionally complete, subsumption can be reduced to concept satisfiability
and \emph{vice versa}: $C\Implies D$ iff $C\And\Not D$ is unsatisfiable and $C$ is satisfiable
iff not $C\Implies A\And\Not A$, where $A$ is an arbitrary concept name.

In \ALCN, number restrictions can be used to restrict the cardinality of the set of fillers of roles
(role successors). For instance, the concept description:
\begin{center}\tt
$\atmost{3}{ \text{child} } \And \all{ \text{child} }{ \text{Female} }$
\end{center}
defines individuals who have at most three daughters and no sons.
Moreover, \ALCNX{\comp} \cite{BS99} allows counting successors of role chains in concept descriptions,
which can be used to express interesting cardinality constraints on the indirect interrelationships
some individuals hold with other objects of the domain.
For example, the \ALCNX{\comp}-concept:
\begin{center}\tt
Man $\!\And \;\atleast{50}{ \text{(friend$\,\comp\,$tel\_number)}  }$
\end{center}
allows us to define men for which the count of different telephone numbers
of their friends amounts at least to fifty.
Notice that such description does not impose further constraints (disregarding obvious ones)
either on the number of friends one may have, or on the number of telephone numbers each friend may
have  (e.g. some friends might have no telephone at all), or even on the fact that some numbers
may be shared by more than one friends (e.g. if husband and wife).
It only gives, for example, a constraint on the minimum size of a phonebook such men need.

The additional role constructs we consider in this paper
further improve the expressiveness of the resulting DLs
and, thus, make them very appealing from an application viewpoint.
For instance, we may use
the \ALCNNX{\comp,\inv{}}-concept:
\begin{center}\tt
Person $\!\And \;\some{ \inv{\text{child}} }{\,\text{Person}}\And\!
\;\atmost{1}{ \text{($\inv{\text{child}}\,\comp\,$child)}  }$
\end{center}
to define persons who are a only child, or the \ALCQX{\comp} concept:
\begin{center}\tt
Woman $\!\And \;\atleastq{3}{ \text{(husband$\,\comp\,$brother)} }{\,\text{Lawyer} }$
\end{center}
to describe women having at least three lawyers as brother-in-law.

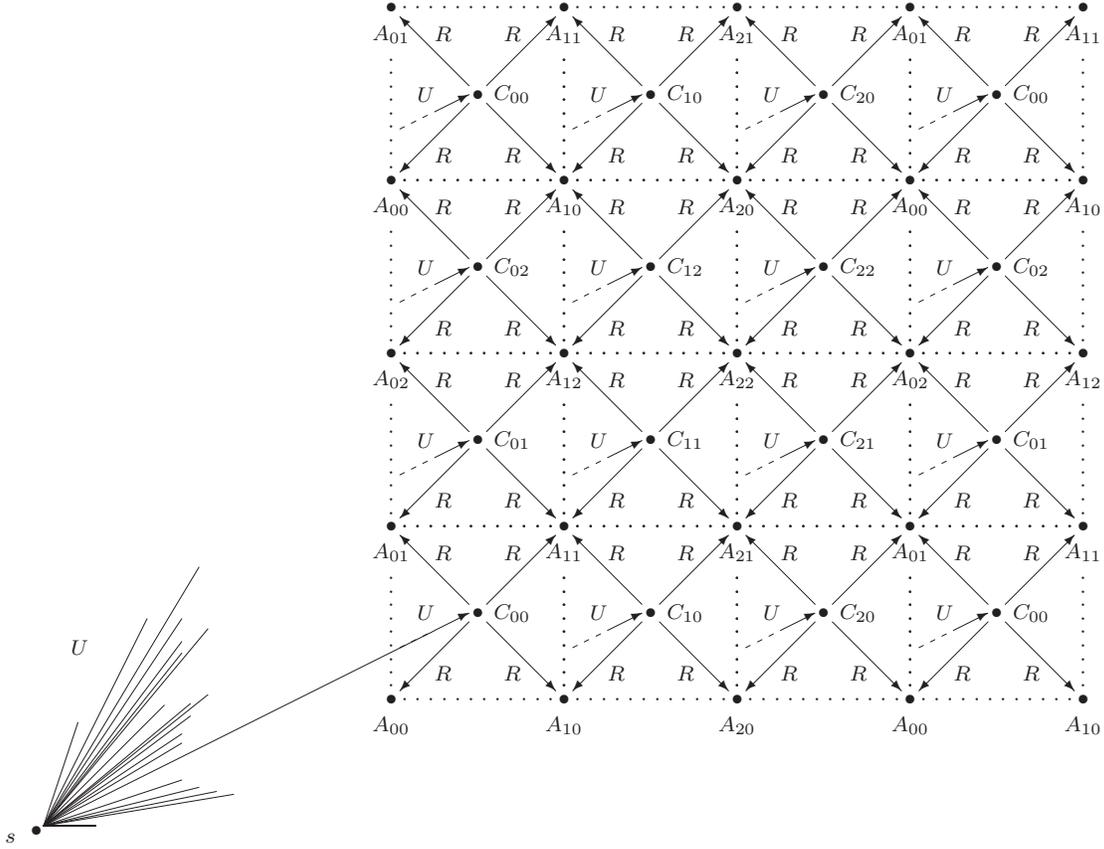
\begin{figure}[tH]
\begin{center}
\begin{scriptsize}
\setlength{\unitlength}{0.23mm}
\begin{picture}(400,500)(0,-75)


\newsavebox{\cell}
\sbox{\cell}{
\xput{50,50}{\bullet}
\put(45,45){\vector(-1,-1){40}}\put(55,55){\vector(1,1){40}}
\put(45,55){\vector(-1,1){40}}\put(55,45){\vector(1,-1){40}}
\dottedline[$\cdot$]{8}(0,70)(0,0)(100,0)(100,70)
\dottedline[$\cdot$]{8}(0,100)(100,100)
\matrixput(0,0)(40,0){2}(0,70){2}{\xput{30,15}{R}}
\put(25,40){\vector(2,1){20}}
\dashline{4}(5,30)(21,38)
\xput{20,50}{U}
}

\matrixput(100,0)(100,0){5}(0,100){5}{\xput{0,0}{\bullet}}
\matrixput(100,0)(100,0){4}(0,100){4}{\usebox{\cell}}

\matrixput(100,0)(300,0){2}(0,300){2}{\xput{0,-15}{A_{00}}\xput{100,-15}{A_{10}}}
\matrixput(100,0)(300,0){1}(0,300){2}{\xput{200,-15}{A_{20}}}
\matrixput(100,100)(300,0){2}(0,300){2}{\xput{0,-15}{A_{01}}\xput{100,-15}{A_{11}}}
\matrixput(100,100)(300,0){1}(0,300){2}{\xput{200,-15}{A_{21}}}
\multiput(100,200)(300,0){2}{\xput{0,-15}{A_{02}}\xput{100,-15}{A_{12}}}
\put(100,200){\xput{200,-15}{A_{22}}}

\matrixput(150,50)(300,0){2}(0,300){2}{\xput{20,0}{C_{00}}}
\multiput(250,50)(0,300){2}{\xput{20,0}{C_{10}}\xput{120,0}{C_{20}}}
\multiput(150,150)(300,0){2}{\xput{20,0}{C_{01}}\xput{20,100}{C_{02}}}
\xput{270,150}{C_{11}}
\xput{370,150}{C_{21}}
\xput{270,250}{C_{12}}
\xput{370,250}{C_{22}}

\xput{-80,30}{U}
\xput{-120,-80}{s}
\xput{-105,-76}{\bullet}
\put(-101,-73){\line(2,1){226}}
\put(-101,-73){\line(1,3){20}}
\put(-101,-73){\line(1,2){60}}
\put(-101,-73){\line(2,3){80}}
\put(-101,-73){\line(3,4){80}}
\put(-101,-73){\line(3,5){90}}
\put(-101,-73){\line(4,5){80}}
\put(-101,-73){\line(5,6){95}}
\put(-101,-73){\line(1,1){70}}
\put(-101,-73){\line(3,2){80}}
\put(-101,-73){\line(4,3){85}}
\put(-101,-73){\line(5,3){80}}
\put(-101,-73){\line(5,4){95}}
\put(-101,-73){\line(6,5){85}}
\put(-101,-73){\line(3,1){80}}
\put(-101,-73){\line(4,1){90}}
\put(-101,-73){\line(5,1){100}}
\put(-101,-73){\line(6,1){110}}
\put(-101,-73){\line(1,0){30}}


\end{picture}
\end{scriptsize}
\end{center}
\caption{\label{grid2} The grid structure used in the \ALCNNX{\comp,^-} undecidability proof.}
\end{figure}

\section{Undecidability of \ALCNNX{\comp,^-}} \label{undec1}
We consider in this Section the extension of $\ALCNX{\comp}$ by inverse roles ($\I$).
Notice that allowing the use of role inversion both in number and in value restrictions,
we obtain a Logic which is a syntactic variant of $\ALCNNX{\comp,^-}$.
Obviously, \ALCNoI concept descriptions are also \ALCNNX{\comp,^-}  concept descriptions.
Conversely, by recursively applying rules $(R\comp S)^- = S^-\comp R^-$
(pushing inverses inwards and eliminating parentheses) and $(R^-)^- = R$,
we can put any $\ALCNNX{\comp,^-}$ complex role expression in the form
$\bar R_1\comp\bar R_2\comp\cdots\comp\bar R_n$,
where each $\bar R_i$ is either an atomic role or the inverse of an atomic role
($\bar R_i\in\set{R_i,R_i^-}$).
Then we can get rid of role composition in value restrictions thanks to the
following equivalences:
\begin{eqnarray*}
  \some{(\bar R_1\comp\bar R_2\comp\cdots\comp\bar R_n)}{C} &\equiv& \some{\bar R_1}{ \some{\bar R_2}{\cdots \some{\bar R_n}{C} } }\\
  \all{(\bar R_1\comp\bar R_2\comp\cdots\comp\bar R_n)}{C} &\equiv& \all{\bar R_1}{ \all{\bar R_2}{\cdots \all{\bar R_n}{C} } }
\end{eqnarray*}
This procedure gives an effective translation of concept descriptions from \ALCNNX{\comp,^-} to {\ALCNoI}.

To show undecidability of \ALCNNX{\comp,^-}, borrowing the proof procedure from \cite{BS99},
we use a reduction of the well-known undecidable domino problem \cite{domino}:
\begin{Definition}
A tiling system ${\cal D} = (D,H,V)$ is given by a non-empty set $D=\set{D_1,\ldots,D_m}$
of domino types, and by horizontal and vertical matching pairs $H\subseteq D\times D$,
$V\subseteq D\times D$. The (unrestricted) domino problem asks for a compatible tiling
of the plane, i.e. a mapping $t: \Z\times\Z \rightarrow D$ such that,
for all $m,n\in \Z$,
\begin{eqnarray*}
  \langle\, t(m,n),t(m+1,n) \,\rangle \in  H 
& \text{and} &
  \langle\, t(m,n),t(m,n+1) \,\rangle \in  V 
\end{eqnarray*}
\end{Definition}

We will show reducibility of the domino problem to \emph{concept satisfiability} in
\ALCNNX{\comp,^-}. In particular, we show how a given tiling system $\cal D$
can be translated into a concept $E_{\cal D}$ which is satisfiable
iff $\cal D$ allows for a compatible tiling.
Following the same lines of undecidability proofs in \cite{BS99}, such translation
can be split into three subtasks which can be described as follows:
\begin{description}
  \item[Grid specification] It must be possible to represent a ``square'' of $\Z\times\Z$,
which consists of points $(m,n),(m+1,n),(m,n+1)$ and $(m+1,n+1)$, in order to yield a
complete covering of the plane via a repeating regular grid structure.
The idea is to introduce concepts to represent the grid points and role expressions to represent
the $x$- and $y$-successor relations. 
  \item[Local compatibility] It must be possible to express that a tiling is locally compatible, that is
that the $x$-successor and the $y$-successor of a point have an admissible domino type.
The idea is to associate each domino type $D_i$ with an atomic concept $D_i$, and to express
the horizontal and vertical matching conditions via value restrictions.
  \item[Total reachability] It must be possible to impose the above local conditions on all points
in $\Z\times\Z$. This can be achieved by constructing a ``universal'' role and a ``start''
individual such that every grid point can be reached from the start individual. The local
compatibility conditions can then be globally imposed via value restrictions.
\end{description}

The grid structure that we will use to tile the plane is shown in Fig.\ref{grid2}.
In particular, in addition to grid points, we also consider ``centers'' of grid squares,
which are connected to grid square vertices by means of a role named $R$.
All grid cell centers are instances
of the $C$ concept, whereas grid points are instances of the $A$ concept.
We introduce nine different (disjoint) types of grid centers via the concepts $C_{ij}$ ($0\leq i,j\leq 2)$
and nine different types of (disjoint) grid points via the concepts $A_{ij}$ ($0\leq i,j\leq 2)$, as follows:
\begin{eqnarray*}
C & := &  \biggOr{0\leq i,j\leq 2}
\bigg(
       C_{ij}\And ( \biggAnd{\substack{0\leq k,\ell\leq 2\\ (i,j)\neq (k,\ell)}} \Not C_{k\ell} )
\bigg)\\ 
A & := &  \biggOr{0\leq i,j\leq 2}
\bigg(
       A_{ij}\And ( \biggAnd{\substack{0\leq k,\ell\leq 2\\ (i,j)\neq (k,\ell)}} \Not A_{k\ell} )
\bigg) \And \Not C
\end{eqnarray*}

\begin{table}
\begin{equation*}
  \begin{array}{|llc|}\hline
    R\text{-successors} & (R\comp R^-)\text{-successors} & C_{ij}\text{-types}\\ \hline\hline
    A_{ij} & C_{ij} & \ding{202} \\
           & C_{i\oplus 2,j}& \ding{203} \\
           & C_{i, j\oplus 2}& \ding{204} \\
           & C_{i\oplus 2,j\oplus 2} & \ding{205} \\ \cline{2-2}
    A_{i\oplus 1,j}
           & C_{i\oplus 1,j} & \ding{206}  \\
           & C_{i\oplus 1\oplus 2,j} = C_{ij} & \ding{192} \\
           & C_{i\oplus 1, j\oplus 2} & \ding{207} \\
           & C_{i\oplus 1\oplus 2,j\oplus 2} = C_{i,j\oplus 2} & \ding{194} \\ \cline{2-2}
    A_{i,j\oplus 1}
           & C_{i,j\oplus 1} & \ding{208} \\
           & C_{i\oplus 2,j\oplus 1} & \ding{209} \\
           & C_{i, j\oplus 1\oplus 2} = C_{ij} & \ding{192} \\
           & C_{i\oplus 2,j\oplus 1\oplus 2} = C_{i\oplus 2,j} & \ding{193} \\  \cline{2-2}
    A_{i\oplus 1,j\oplus 1}
           & C_{i\oplus 1,j\oplus 1} & \ding{210} \\
           & C_{i\oplus 1\oplus 2,j\oplus 1} = C_{i,j\oplus 1} & \ding{198} \\
           & C_{i\oplus 1, j\oplus 1\oplus 2} = C_{i\oplus 1,j} & \ding{196} \\
           & C_{i\oplus 1\oplus 2,j\oplus 1\oplus 2} = C_{ij} & \ding{192} \\ \hline
  \end{array}
\end{equation*}
\caption{\label{tab1}
Types of the $R$- and $(R\comp R^-)$-successors of a $C_{ij}$-type grid center.
In the last column, numbers on black ground mark different $C_{ij}$-types the first time
they are met from the top of the table, whereas numbers on white ground refer to
$C_{ij}$-types that have been met before.}
\end{table}

\noindent{\bf Grid specification} can then be accomplished by means of the $C_\boxplus$
and $A_\boxplus$ concepts which follow:
\begin{eqnarray*}
C_{\boxplus} & := & C\And \atmost{4}{R}
                     \And \all{R}{A_{\boxplus}} 
                     \And \atmost{9}{R\comp R^-} \And \\ 
&& \biggAnd{0\leq i,j\leq 2}
\big(
       C_{ij}\Rightarrow (
 \some{R}{A_{ij}}\And\some{R}{A_{i\oplus 1,j}} \And 
   \some{R}{A_{i,j\oplus 1}}\And \some{R}{A_{i\oplus 1,j\oplus 1}} ) \big) \\
A_{\boxplus} & := & A\And
\biggAnd{0\leq i,j\leq 2}
\big(
       A_{ij}\Rightarrow (
 \some{R^-}{C_{ij}}\And\some{R^-}{C_{i\oplus 2,j}} \And 
   \some{R^-}{C_{i,j\oplus 2}}\And \some{R^-}{C_{i\oplus 2,j\oplus 2}} ) \big) \\
\end{eqnarray*}
where $A\Rightarrow B$ is a shorthand for $\Not A\Or B$
and $a\oplus b=(a+b)\bmod 3$.

Some relevant constraints that are imposed by these concept descriptions
on their models are studied in the Lemmata and Corollaries which follow.
\begin{lemma}\label{lemma10}
Let $c$ be an instance of $C_{\boxplus}$. Then it has at most one $R$-successor
in each of the nine $A_{k\ell}$ concept extensions.
\end{lemma}
\textbf{Proof}
More precisely, $c$ has exactly one $R$-successor
in the extension of each of the four $A_{k\ell}$ concepts it is connected to by $R$
(e.g. if w.l.o.g. $c\in\Int{(C_{ij})}$ then $c$ has exactly one $R$-successor
in the extension of $A_{ij}$, $A_{i\oplus 1,j}$, $A_{i,j\oplus 1}$, $A_{i\oplus 1,j\oplus 1}$
and no $R$-successor in any of the remaining five partitions of the extension of $A$).
This follows from the fact that the nine $A_{k\ell}$ concepts are disjoint
and $c$ has a total number of at most four $R$-successors.
\QED

\begin{lemma}\label{lemma11}
Let $c$ be an instance of $C_{\boxplus}$. Then it has exactly one $(R\comp R^-)$-successor
in each of the nine $C_{k\ell}$ concept extensions.
\end{lemma}
\textbf{Proof}
Since $c$ is an instance of $C$, it belongs to the extension of exactly one of the nine $C_{k\ell}$ concepts.
W.l.o.g. let us assume $c\in\Int{(C_{ij})}$. Hence,
owing to the $C_{\boxplus}$ and $A_{\boxplus}$ definitions and Lemma~\ref{lemma10},
$c$ has surely $R$-
and $(R\comp R^-)$-successors as shown in Tab. \ref{tab1}. In particular, $c$
has $(R\comp R^-)$-successors in each of the nine $C_{k\ell}$ concept extensions.
Since all the $C_{k\ell}$ extensions are disjoint and $c$ has a total of at most nine
$(R\comp R^-)$-successors, this means that $c$ has exactly one $(R\comp R^-)$-successor in each
of the nine $C_{k\ell}$ concept extensions (being $c$ itself its unique $(R\comp R^-)$-successor
in $C_{ij}$). \QED

\begin{corollary}\label{coro10}
Let $a$ be an instance of $A_{\boxplus}$. Then (1) all its $(R^-)$-successors are
instances of $C_{\boxplus}$ and (2) it has at most one $(R^-)$-successor
in each of the nine $C_{k\ell}$ concept extensions.
\end{corollary}
\textbf{Proof} It is an immediate consequence of Lemma~\ref{lemma11}.
W.l.o.g. assume $a\in\Int{(A_{ij})}$.
(1) If $a$ had an  $(R^-)$-successor $o\not\in\Int{(C_{\boxplus})}$,
then any of the four $(R^-)$-successors of $a$ in $C_{\boxplus}$ would have
at least ten $(R\comp R^-)$-successors (it has nine $(R\comp R^-)$-successors in
$C_{\boxplus}$ by Lemma~\ref{lemma11} plus $o$) and, thus, would violate
the $C_{\boxplus}$ definition.
(2) If $a$ had, for instance, two
distinct $(R^-)$-successors in $C_{k\ell}$ (i.e. $\exists c,c'\in\Int{(C_{k\ell})}$,
$c\neq c'$, with $(c,a)\in\Int R$, $(c',a)\in\Int R$), then
$c$ would have two distinct $(R\comp R^-)$-successors
in $C_{k\ell}$, $c'$ and itself, contradicting Lemma~\ref{lemma11}.

More precisely, $a$ has exactly one $(R^-)$-successor
in the extension of each of the four $C_{k\ell}$ concepts it is connected to by $R^-$.
\QED

\begin{corollary}\label{coro11}
Let $a$ be an instance of $A_{\boxplus}$. Then it has exactly one $(R^-\comp R)$-successor
in each of the nine $A_{k\ell}$ concept extensions.
\end{corollary}
\textbf{Proof}
W.l.o.g. assume $a\in\Int{(A_{ij})}$. We show that if the thesis is false
we come up with a contradiction. To this end, we must distinguish three cases.
First of all, we can exclude $a$ has another $(R^-\comp R)$-successor, say $a'$,
in $A_{ij}$: if this happened, each of the four $(R^-)$-successors of $a$
(e.g. $c\in\Int{(C_{ij})}$) would have two distinct $R$-successors ($a$ and $a'$)
in $A_{ij}$, thus violating Lemma~\ref{lemma10}.
Second, we can also exclude $a$ has two distinct $(R^-\comp R)$-successors in  $A_{k\ell}$,
say $a'$ and $a''$, which can be reached through a common $(R^-)$-successor $c$
(e.g. $c\in\Int{(C_{ij})}$): if this happened, $c$ would have two distinct $R$-successors
($a$ and $a'$) in $A_{k\ell}$, against Lemma~\ref{lemma10} again.
In the third and last case, we must consider $a$ having two distinct $(R^-\comp R)$-successors
in  $A_{k\ell}$, say $a'$ and $a''$, which can be reached through distinct $(R^-)$-successors of $a$.
W.l.o.g. we may assume such $(R^-)$-successors of $a$ in $C_{ij}$ and $C_{i,j\oplus 2}$,
and $a',a''\in\Int{(A_{i\oplus 1,j})}$.
%
%
Hence we must have that $\exists c_0\in\Int{(C_{ij})}$ with $\set{(c_0,a),(c_0,a')}\subseteq\Int R$
and  $\exists c\in\Int{(C_{i,j\oplus 2})}$ with $\set{(c,a),(c,a'')}\subseteq\Int R$.
We then consider the application of Lemma~\ref{lemma11} from $c_0\in\Int{(C_{ij})}$.
By construction, $c_0$ has $c$ as $(R\comp R^-)$-successor through the path
passing from $a\in\Int{(A_{ij})}$. Owing to Lemma~\ref{lemma11}, also the
$(R\comp R^-)$ path passing from $a'\in\Int{(A_{i\oplus 1,j})}$
(the path exists, as $a'$ has an $(R^-)$-successor in $\Int{(C_{i,j\oplus 2})}$)
must lead to $c$ and, thus, $(c,a')\in\Int R$.
But this contradicts Lemma~\ref{lemma10}, as $c$ would have two distinct $R$-successors
($a'$ and $a''$) in $\Int{(A_{i\oplus 1,j})}$. \QED

Hence, we will interpret instances of $C_{\boxplus}$ as grid centers
and instances of $A_{\boxplus}$ as grid points.
In particular, nine different types of grid cells can be defined according to the type of their center:
an $(i,j)$-type grid cell has a $C_{ij}$-type center, while its lower left, lower right,
upper left and upper right vertices can be defined, respectively, as the instances of the
$A_{ij}$, $A_{i\oplus 1,j}$, $A_{i,j\oplus 1}$ and $A_{i\oplus 1,j\oplus 1}$ concepts which are connected
to the center via $R$ (according to the $C_{\boxplus}$ definition).
Therefore, the $x$- and $y$-successor relations on the grid can be defined
by means of the $(R^-\comp R)$-paths connecting an $A_{ij}$-type grid point
with an $A_{i\oplus 1,j}$-type and an $A_{i,j\oplus 1}$-type grid points, respectively.
Such successors always exist and are uniquely defined, owing to Corollary~\ref{coro11}.

In a similar way, Corollary~\ref{coro11} also allows us to uniquely define the
$x$- and $y$-predecessors relations on the grid, by means of the
$(R^-\comp R)$-paths connecting an $A_{ij}$-type grid point
with an $A_{i\oplus 2,j}$-type and an $A_{i,j\oplus 2}$-type grid points, respectively
(cf. $(a-1)\mod 3=(a+2)\mod 3$).

\begin{lemma}[Grid Closure]\label{lemma12}
For each grid point, the $(x\comp y)$-
and $(y\comp x)$-successors are uniquely defined and coincide.
\end{lemma}
\textbf{Proof}
We can assume the grid point  to represent the point $(m,n)\in\Z\times\Z$ and call it $p_{(m,n)}$.
W.l.o.g. we can further assume  $p_{(m,n)}$ to be the bottom left vertex of an $(i,j)$-type grid cell.
Therefore, $p_{(m,n)}$ is an instance of $A_{ij}$ and is an $R$-successor of the grid cell center,
say $c_{(m,n)}$, which is an instance of $C_{ij}$.
The $x$-successor of  $p_{(m,n)}$, say  $p_{(m+1,n)}$, is the $R$-successor of  $c_{(m,n)}$
in $A_{i\oplus 1,j}$ (by construction, it is an $(R^-\comp R)$-successor of  $p_{(m,n)}$ and is
unique by Corollary~\ref{coro11}).
Analogously, the $y$-successor of  $p_{(m,n)}$, say  $p_{(m,n+1)}$, is the $R$-successor of  $c_{(m,n)}$
in $A_{i,j\oplus 1}$.
According to the $C_{\boxplus}$ definition,  $c_{(m,n)}$ has also a fourth $R$-successor,
say $\bar p$, in $A_{i\oplus 1,j\oplus 1}$.
We consider now the  $(x\comp y)$-successor of $p_{(m,n)}$, that is the $y$-successor of $p_{(m+1,n)}$,
and call it $p'_{(m+1,n+1)}$. Owing to the $y$-successor definition,  $p'_{(m+1,n+1)}$
must be an instance of $A_{i\oplus 1,j\oplus 1}$ connected to  $p_{(m+1,n)}$ via $(R^-\comp R)$.
However, both $\bar p$ and $p'_{(m+1,n+1)}$ are, by construction, $(R^-\comp R)$-successors
of $p_{(m+1,n)}$ in $A_{i\oplus 1,j\oplus 1}$ and, thus, they must coincide thanks to Corollary~\ref{coro11}.
Analogously,  the  $(y\comp x)$-successor of $p_{(m,n)}$, that is the $x$-successor of $p_{(m,n+1)}$,
say $p''_{(m+1,n+1)}$, must be an instance of $A_{i\oplus 1,j\oplus 1}$ connected
to  $p_{(m,n+1)}$ via $(R^-\comp R)$. Thence, Corollary~\ref{coro11} ensures that $\bar p$
and  $p''_{(m+1,n+1)}$ coincide, as they are both  $(R^-\comp R)$-successors
of $p_{(m,n+1)}$ in $A_{i\oplus 1,j\oplus 1}$.
Hence, $\bar p$ is the common $(x\comp y)$-
and $(y\comp x)$-successor of $p_{(m,n)}$ on the grid, that can be called $p_{(m+1,n+1)}$
to represent the point $(m+1,n+1)$ of the plane. \QED

\vskip 5mm
\noindent{\bf Local compatibility} is easily achieved by enforcing grid centers
to be instances of a $C_{\cal D}$ concept defined as follows:
\begin{align*}
C_{\cal D} \; := \; &
\all{R}{ \bigg( \biggOr{1\leq k\leq m} \big(
                       D_k\And ( \biggAnd{\substack{1\leq \ell\leq m\\ k\neq \ell}} \Not D_\ell ) \big)\,\bigg) }\And
  \biggAnd{0\leq i,j\leq 2}
\bigg(\, C_{ij}\Rightarrow \biggAnd{1\leq k\leq m} \some{R}{ (A_{ij}\And D_k)} \\
   & \quad\quad  \Rightarrow   \big( \, \some{R}{(A_{i\oplus 1,j}\And (\bigOr{(D_k,D_\ell)\in H} D_\ell))} \And
              \some{R}{(A_{i,j\oplus 1}\And (\bigOr{(D_k,D_\ell)\in V} D_\ell))} 
\, \big) \,\bigg)
\end{align*}
Each domino type $D_k$ is associated to an atomic concept with the same name. The value restriction
in the first conjunct of $C_{\cal D}$ forces grid points to have a domino type.
The second conjunct uses the definition of the $x$- and $y$-successors for the bottom left vertex
of an $(i,j)$-type cell to enforce horizontal and vertical matching conditions via value restrictions.

\vskip 5mm
\noindent\textbf{Total Reachability}
will be achieved by constructing a ``start'' individual ($s$)
and two ``universal'' roles: the former ($U$) which connects $s$ to every grid center
and the latter ($U\comp R$) which connects $s$ to every grid point  (see Fig.~\ref{grid2}).
The Lemmata and Corollaries which follow will justify the correctness of our construction.

\begin{lemma}\label{lemma1}
Let $s$ be an instance of
\begin{equation*}
  D\;:=\;\exists U\comp R\And \atmost{1}{(U\comp R)\comp (U\comp R)^-}
\And\Not\exists R^-\And\Not\exists U^-\And\all{U}{\Not\exists R^-}
\end{equation*}
in a given interpretation $\I$.
Then any $(U\comp R)$-successor $x$ of $s$ in $\I$
($D$ ensures that there is at least one) has $s$ as its
\emph{unique} $(U\comp R)$-predecessor.
\end{lemma}
\textbf{Proof} Assume $s\in\Int D$ and $x$ is a $(U\comp R)$-successor of $s$,
that is $\exists o\in\Int\Delta$ such that $(s,o)\in\Int U$, $(o,x)\in\Int R$,
with $o\neq s$ (as $s\not\in\Int{(\exists U^-)}$),
$x\neq o$ (as $s\in\Int{(\all{U}{\Not\exists R^-})}$ and, thus, $o\not\in\Int{(\exists R^-)}$)
and $s\neq x$ (as $s\not\in\Int{(\exists R^-)}$).
If there were $s'\in\Int\Delta$, $s'\neq s$, such that $s'$ is a $(U\comp R)$-predecessor
of $x$ (i.e. $\exists o'\in\Int\Delta$ such that $(s',o')\in\Int U$, $(o',x)\in\Int R$),
then $s$ and $s'$ would be both $(R^-\comp U^-)=(U\comp R)^-$-successors of $x$ in $\I$
and, thus, both $(U\comp R)\comp (U\comp R)^-$-successors of $s$ in $\I$.
Hence we should have $s\not\in\Int D$, against the hypothesis. \QED

\begin{corollary}\label{corol1}
Under the hypothesis of Lemma \ref{lemma1},
any $U$-successor of $s$ in $\I$ has $s$ as its \emph{unique} $U$-predecessor.
\end{corollary}
\begin{corollary}\label{corol2}
Under the hypothesis of Lemma \ref{lemma1},
let $s$ be an instance of
\begin{equation*}
  D'\;:=\;D\And \all{U}{\all{R}{\all{R^-}{\exists U^-}}} \And\Not\exists R
\end{equation*}
in a given interpretation $\I$.
Then any $(U\comp R\comp R^-)$-successor $y$ of $s$ in $\I$
($D$ ensures that there is at least one) is a $U$-successor of $s$ in $\I$
and has $s$ as its \emph{unique} $U$-predecessor.
\end{corollary}
\textbf{Proof} Let $y\in\Int \Delta$ be a generic $(U\comp R\comp R^-)$-successor of $s$
in $\I$, that is $\exists o,x\in\Int\Delta$ such that $(s,o)\in\Int U$, $(o,x)\in\Int R$, $(y,x)\in\Int R$
(we may assume $y\neq o$, as the Corollary is trivially true for $o$),
with $s\neq y$ (as $s\not\in\Int{(\exists R)}$).
Since $s\in\Int{(\all{U}{\all{R}{\all{R^-}{\exists U^-}}})}$, $y\in\Int{(\exists U^-)}$,
that is $\exists s'\in\Int\Delta$ such that $(s',y)\in\Int U$.
Notice that both $s$ and $s'$ have, by construction, $x$ as $(U\comp R)$-successor.
Since $s\in\Int D$, thanks to Lemma \ref{lemma1}, $s$ and $s'$ must coincide.
Hence $y$ is a $U$-successor of $s$, which is also its unique $U$-predecessor
by Corollary \ref{corol1}. \QED

\begin{lemma}[Plane Covering and Compatible Tiling] \label{lemma2}
Let $s$ be an instance of
\begin{eqnarray*}
E_{\cal D} &:=&
\exists U\comp R \And \atmost{1}{(U\comp R)\comp (U \comp R)^-}
\And \Not\exists R \And \Not\exists R^- \And\Not\exists U^- \And \\
 & & \all{U}{\all{R}{\all{R^-}{\exists U^-}}} \And
 \all{U}{( C_{\boxplus}\And 
C_{\cal D} \And \Not\exists R^- )} 
\end{eqnarray*}
in a given interpretation $\I$. Then, for the grid that tiles the plane $\Z\times\Z$,
any grid center can be reached from $s$ via $U$,
any grid point can be reached from $s$ via $U\comp R$ and local tiling conditions
are imposed on all grid points (yielding a compatible tiling of the plane).
\end{lemma}
\textbf{Proof}
Let us consider a grid center connected to $s\in\Int{(E_{\cal D})}$
via $U$ ($E_{\cal D}\Implies\exists U\comp R\And\all{U}{C_{\boxplus}}$ ensures that
there is at least one).
W.l.o.g. we can assume  it to be the center of an $(i,j)$-type cell and call it $c_{(0,0)}$
($c_{(0,0)}\in\Int{(C_{ij})}$). We can also call $p_{(0,0)}$ the bottom left vertex
of this grid cell ($p_{(0,0)}\in\Int{(A_{ij})}$) and let it represent the origin $(0,0)$ of $\Z\times\Z$.
We can now consider the $x$- and $y$-successors of $p_{(0,0)}$, say $p_{(1,0)}$ and $p_{(0,1)}$, respectively.
By construction, we have $p_{(1,0)}\in\Int{(A_{i\oplus 1,j})}$, $p_{(0,1)}\in\Int{(A_{i,j\oplus 1})}$;
moreover, either $p_{(0,0)}$, $p_{(1,0)}$ and $p_{(0,1)}$ are $R$-successors of $c_{(0,0)}$
and, thus, $(U\comp R)$-successors of $s$.
In the $(i,j)$-type grid cell centered on $c_{(0,0)}$, $p_{(1,0)}$ and $p_{(0,1)}$ are the
bottom right and top left vertices,  but they are also the  bottom left vertices
of the two grid cells adjacent to the right and to the top, respectively.
In particular, $p_{(1,0)}$ and $p_{(0,1)}$ are the bottom left vertices of an $(i\oplus 1,j)$-
and an $(i,j\oplus 1)$-type grid cells, whose centers we can call $c_{(1,0)}$ and $c_{(0,1)}$,
respectively (the existence and uniqueness of these cells and their centers is ensured
by Lemma~\ref{lemma11}). Obviously, $c_{(1,0)}$ and $c_{(0,1)}$ are
($(R\comp R^-)$-successors of $c_{(0,0)}$ and) $(R^-)$-successors
of  $p_{(1,0)}$ and $p_{(0,1)}$, respectively. Therefore, they are
$(U\comp R\comp R^-)$-successors of the start individual $s$ and, thanks to Corollary~\ref{corol2}
(as $E_{\cal D}\Implies D'$), they are also $U$-successors of $s$.

Using the $x$- and $y$-predecessor definitions, we can easily see that the same holds for
$c_{(-1,0)}$ and $c_{(0,-1)}$ grid centers. In any case, we can repeat the argument at will,
starting with $c_{(1,0)}$, $c_{(0,1)}$, $c_{(-1,0)}$ and $c_{(0,-1)}$ in place of $c_{(0,0)}$,
and show that the center of any grid cell on the plane can be reached from $s$ via $U$.
Hence, all grid points can be reached from $s$ via $U\comp R$ and local tiling conditions
are imposed on all of them by value restrictions (as $E_{\cal D}\Implies \all{U}{C_{\cal D}}$).
\QED

\vskip 5mm
Thanks to Lemma~\ref{lemma2}, it is easy to see that
a tiling system $\cal D$ has a compatible tiling
iff concept $E_{\cal D}$ is satisfiable (i.e. there is an interpretation $\I$
such that $\Int{(E_{\cal D})} \neq \emptyset$).
\begin{theorem}\label{th1}
Satisfiability (and, thus, subsumption) of concepts
is undecidable for \ALCNNX{\comp,^-} (and \ALCQX{\comp,^-}).
\end{theorem}


\section{Decidability of \ALCQX{\comp}} \label{dec2}

We will show in this Section how an effective decision procedure for \ALCQX{\comp}-concept
satisfiability can be provided as a tableau-based algorithm.
To this end, we consider \ALCQX{\comp}-concept descriptions in Negation Normal Form (NNF \cite{ALC}),
where the negation sign is allowed to appear before atomic concepts only.
In fact, \ALCQX{\comp}-concept descriptions can be transformed into NNF in linear time
via application of the same rules which can be used for \ALCQ (pushing negations inwards):
\begin{equation*}
\begin{array}{cccc}
 \Not \atmostq{n}{R}{C} \;=\; \atleastq{n+1}{R}{C} &&&
 \Not \atleastq{n}{R}{C} \;=\; \atmostq{n-1}{R}{C} \quad (\Bot \text{ if }n=0)\\
 \Not \some{R}{C} \;=\; \all{R}{\Not C} &&&
 \Not \all{R}{C} \;=\; \some{R}{\Not C} \\
\end{array}
\end{equation*}
in addition to the absorption rule for double negations and De Morgan's laws for $\And$ and $\Or$.
Obviously, unqualified number restrictions are treated as particular cases of qualified restrictions (with $C=\Top$).
We can further make use of the rules:
\begin{eqnarray*}
\begin{array}{cccc}
\some{R}{C} \;=\; \atleastq{1}{R}{C} &&&
\all{R}{C} \;=\; \atmostq{0}{R}{\Not C} 
\end{array}
\end{eqnarray*}
to get rid of (existential and) value restrictions. We define the concept descriptions
obtained in this way as in NNF$^{\Join}$ and denote the NNF$^{\Join}$ of the
\ALCQX{\comp}-concept description $\Not C$ as $\sim C$.
We will use the symbol $\Join$ in number restrictions $\atbothq{n}{R}{C}$ as a placeholder
for either $\geq$ or $\leq$.

The Tableau algorithm we are going to introduce
manipulates, as basic data structures, ABox assertions
involving domain individuals.
In fact, our algorithm is a simple extension
of the tableau-based algorithm to decide \ALCNX{\comp}-concept satisfiability presented by
Baader and Sattler in \cite{BS99}. The extension is based on the modification
of the transformation rules for number restrictions ($\geq$- and $\leq$-rules)
to take into account the ``qualifying'' conditions
and on the introduction of a so-called \emph{choose} rule (called \choos-rule here),
which makes sure that all ``relevant'' concepts that are implicitly satisfied by an individual
are made explicit in the ABox.
Basically, the proposed extension is similar to the one which extends
the tableau-based \ALCN satisfiability algorithm \cite{ALCN1,ALCN2} to
an \ALCQ satisfiability algorithm \cite{HB91,BS01}.

\begin{definition}
Let $\NI$ be a set of individual names. An \emph{ABox} \A\ is a finite set of assertions of the form $C(a)$
--\emph{concept assertion}-- or $R(a,b)$ --\emph{role assertion}-- where $C$ is a concept description, $R$ a role
name, and $a,b$ are individual names.
An interpretation $\I$, which additionally assigns elements $a^\I\in\dom$ to individual names $a$,
is a \emph{model} of an ABox \A\ iff $a^\I\in C^\I$ (resp. $(a^\I,b^\I)\in R^\I$) for all
assertions $C(a)$ (resp. $R(a,b)$) in \A.
The ABox \A\ is consistent iff it has a model. The individual $a$ is an instance of the description $C$ w.r.t.
\A\ iff $a^\I\in C^\I$ holds for all models $\I$ of \A.
We also consider in a ABox \emph{inequality assertions} of the form
$a\neq b$, with the obvious semantics that an interpretation $\I$ satisfies $a\neq b$, iff
$a^\I \neq b^\I$. Inequality assertions are assumed to be symmetric, that is saying that
$\set{a\neq b}\subseteq \A$ is the same as saying $\set{b\neq a}\subseteq\A$.
\end{definition}
Sometimes in the DL field, a \emph{unique name assumption} is made in works concerning reasoning with individuals,
that is the mapping $\pi:\NI\rightarrow\dom$ from individual names to domain elements is required
to be injective. We dispense from this requirement as it has no effect for the \ALC extensions studied here
and the explicitly introduced inequality assertions can be used to enforce the uniqueness of names anyway.

\begin{definition}
The individual $y$ is a $(R_1\comp R_2\comp\cdots\comp R_m)$-\emph{successor} of $x$ in \A\
iff $\exists y_2 y_3 \ldots y_m$ variables in \A\ such that
$\set{ R_k(y_k,y_{k+1}) \mid 2\leq k\leq m-1 }\cup\set{ R_1(x,y_2),R_m(y_m,y) } \subseteq\A$. 
\end{definition}

\begin{definition}
An ABox \A\ contains a \emph{clash} iff, for an individual name $x\in\NI$, one of the two situations below occurs:
\begin{itemize}
\item $\set{A(x),\Not A(x)}\subseteq \A$, for a concept name $A\in\NC$;
\item $(\atmostq{n}{R_1\comp\cdots\comp R_m}{C})(x)\in\A$ and
$x$ has $p$ $(R_1\comp\cdots\comp R_m)$-successors $y_1,\ldots,y_p$ with $p>n$
such that $\set{C(y_i)\mid 1\leq i\leq p}\cup\set{y_i\neq y_j\mid 1\leq i<j\leq p}\subseteq \A$,
for role names $\set{R_1,\ldots,R_m}\subseteq\NR$, a concept description $C$ and an integer $n\geq 0$.
\end{itemize}
\end{definition}

To test the satisfiability of an \ALCQX{\comp} concept $C$ in NNF$^{\Join}$, the proposed \ALCQX{\comp}-algorithm
works as follows. Starting from the initial ABox  $\set{C_0(x_0)}$, it applies the \emph{completion rules}
in Fig.~\ref{fig:tabl1}, which modify the ABox. It stops when no rule is applicable (when a clash is generated,
the algorithm does not immediately stops but it always generate a complete ABox).
An ABox \A\ is called \emph{complete} iff none of the completion rules is any longer applicable.
The algorithm answers ``$C$ is satisfiable'' iff a complete
and clash-free ABox has been generated.
The \ALCQX{\comp}-algorithm is non-deterministic, due to the $\Or$-, $\leq$- and \choos-rules
(for instance, the $\Or$-rule non-deterministically chooses which disjunct to add for a disjunctive concept).

\begin{figure}
\hrule
\begin{eqnarray*}
\And\text{-rule: } & \text{\bf if}\quad 1. & (C_1\And C_2)(x)\in\A \text{ and} \\
                   &        \;\;\;\quad 2. & \set{ C_1(x), C_2(x) }\not\subseteq \A \\
                   & \text{\bf then} & \A':=\A\cup\set{ C_1(x), C_2(x) } \\
 \Or\text{-rule: } & \text{\bf if}\quad 1. & (C_1\Or C_2)(x)\in\A \text{ and} \\
                   &        \;\;\;\quad 2. & \set{ C_1(x), C_2(x) }\cap \A=\emptyset \\
                   & \text{\bf then} & \A':=\A\cup\set{ D(x) } \text{ for some } D\in\set{ C_1,C_2 }\\
\geq\text{-rule: } & \text{\bf if}\quad 1. & (\atleastq{n}{R_1\comp\cdots\comp R_m}{C})(x)\in\A \text{ and} \\
                   &        \;\;\;\quad 2. & x \text{ has exactly $p$ $(R_1\comp\cdots\comp R_m)$-successors $y_1,\ldots,y_p$ with }p<n\\
                                    && \text{such that } \set{C(y_i)\mid 1\leq i\leq p}\cup\set{y_i\neq y_j\mid 1\leq i<j\leq p}\subseteq \A\\
                   & \text{\bf then} & \A':=\A\cup\set{ R_1(x,z_{i2}),R_2(z_{i2},z_{i3}),\ldots,R_m(z_{im},z_i),C(z_i)\mid 1\leq i\leq n-p }\\
                                 &&\qquad\quad\;\;    \cup\set{ z_i\neq z_j\mid 1\leq i<j\leq n-p}
                                                      \cup\set{ y_i\neq z_j\mid 1\leq i\leq p, 1\leq j\leq n-p}\\
                                    && \text{where $z_{ik},z_i$ (for $1\leq i\leq n-p,2\leq k\leq m$) are $m(n-p)$ fresh variables } \\
\leq\text{-rule: } & \text{\bf if}\quad 1. & (\atmostq{n}{R_1\comp\cdots\comp R_m}{C})(x)\in\A \text{ and} \\
                   &        \;\;\;\quad 2. & x \text{ has more than $n$ $(R_1\comp\cdots\comp R_m)$-successors $y_1,\ldots,y_p$ such that}\\
 && \set{ C(y_i)\mid 1\leq i\leq p }\subseteq \A \text{ and } \set{y_i\neq y_j}\cap\A=\emptyset \text{ for some $i,j$ ($1\leq i<j\leq p$), } \\
                   & \text{\bf then} & \text{for some pair }y_i,y_j \,(1\leq i<j\leq p) \text{ such that } \set{y_i\neq y_j}\cap\A=\emptyset \\
&& \A':=[y_i/y_j]\A \text{ (i.e. $\A'$ is obtained 
by replacing each occurrence of $y_i$ by $y_j$)}\\
\text{\choos-rule: } & \text{\bf if}\quad 1. & (\atbothq{n}{R_1\comp\cdots\comp R_m}{C})(x) \in\A \text{ and }\\
                   &        \;\;\;\quad 2. &  y \text{ is an $(R_1\comp\cdots\comp R_m)$-successor of $x$ such that }
                  \set{ C(y), \sim C(y) }\cap \A=\emptyset \\
                   & \text{\bf then} & \A':=\A\cup\set{ D(y) } \text{ for some } D\in\set{ C,\sim C }\\
\end{eqnarray*}
\hrule
\caption{\label{fig:tabl1}
The Completion Rules for \ALCQX{\comp}}
\vspace*{3mm}\hrule
\end{figure}

\begin{lemma}\label{lemmaX}
Let $C_0$ be an  \ALCQX{\comp}-concept in NNF$^{\Join}$, and let \A\ be an ABox obtained by applying
the completion rules to  $\set{C_0(x_0)}$. Then
\begin{enumerate}
  \item For each completion rule $\cal R$ that can be applied to \A\ and for each interpretation $\I$,
the following equivalence holds: $\I$ is a model of \A\ iff $\I$ is a model of the ABox
$\A'$ obtained by applying  $\cal R$.
  \item If \A\ is a complete and clash-free ABox, then \A\ has a model.
  \item If \A\ is complete but contains a clash, then \A\ does not have a model.
  \item The completion algorithm terminates when applied to  $\set{C_0(x_0)}$.
\end{enumerate}
\end{lemma}
As a matter of fact, termination (4) yields that after finitely many steps
we obtain a complete ABox. If $C_0$ is satisfiable,
then  $\set{C_0(x_0)}$ is also satisfiable and, thus, at least one of the
complete ABoxes that the algorithm can generate is satisfiable by (1).
Hence, such an ABox must be clash-free by (3). Conversely, if the application
of the algorithm produces a complete and clash-free ABox \A, then it is
satisfiable by (2) and, owing to (1), this implies that $\set{C_0(x_0)}$ is satisfiable.
Consequently, the algorithm is a decision procedure for satisfiability
of  \ALCQX{\comp}-concepts.

\begin{corollary}
Concept satisfiability (and subsumption) for \ALCQX{\comp} is decidable,
and the Tableau algorithm based on the completion rules in Fig.~\ref{fig:tabl1}
is an effective decision procedure. 
\end{corollary}

\noindent\textbf{Proof of Part 1 of Lemma \ref{lemmaX}}
We consider only the rules concerned with number restrictions
and the \choos-rule, as the proof for the first two rules
is the same as for \ALC.
\begin{description}
\item[3. $\geq$-rule.] Assume that the rule is applied to the constraint
$(\atleastq{n}{R_1\comp\cdots\comp R_m}{C})(x)$ and that its application yields:
\begin{eqnarray*}
 \A'& = & \A\cup\set{ R_1(x,z_{i2}),R_2(z_{i2},z_{i3}),\ldots,R_m(z_{im},z_i),C(z_i)\mid 1\leq i\leq n-p }\\
                                 &&\qquad \cup\set{ z_i\neq z_j\mid 1\leq i<j\leq n-p}
                                          \cup\set{ y_i\neq z_j\mid 1\leq i\leq p, 1\leq j\leq n-p}
\end{eqnarray*}
Since \A\ is a subset of $\A'$, any model of  $\A'$ is also a model of \A.
Conversely, assume that $\I$ is a model of \A. On the one hand, since $\I$ satisfies
$(\atleastq{n}{R_1\comp\cdots\comp R_m}{C})(x)$, $x^\I$ has at least $n$
$(R_1\comp\cdots\comp R_m)$-successors in $\I$ which are instances of $C$.
On the other hand, since the  $\geq$-rule
is applicable to $(\atleastq{n}{R_1\comp\cdots\comp R_m}{C})(x)$, $x$ has exactly
$p$ $(R_1\comp\cdots\comp R_m)$-successors $y_1,\ldots,y_p$, with $p<n$, which are instances of $C$ in $\A$.
Thus, there exists $n-p$ $(R_1\comp\cdots\comp R_m)$-successors
$b_1,\ldots,b_{n-p}$ of $x^\I$ in $\I$
such that $b_i\in C^\I$ and $b_i\neq y_j$ for all $i,j$ ($1\leq i\leq n-p,1\leq j\leq p$).
For all $i$ ($1\leq i\leq n-p$), let $\set{b_{i2},\ldots, b_{im}}\subseteq\dom$
be such $(x^\I,b_{i2})\in R_1^\I$, $(b_{i2},b_{i3})\in R_2^\I,\ldots,$
$(b_{im},b_i)\in R_m^\I$. We define the interpretation of the new variables
added by the $\geq$-rule as
$z_{i2}^\I=b_{i2},\ldots,z_{im}^\I=b_{im}$, and $z_i^\I=b_i$ ($1\leq i\leq n-p$).
Obviously, $\I$ satisfies $\A'$.
\item [4. $\leq$-rule.] Assume that the rule is applied to the constraint
$(\atmostq{n}{R_1\comp\cdots\comp R_m}{C})(x)\in \A$ and let $\I$ be a model of \A.
On the one hand, since the rule is applicable, $x$ has more than
$n$ $(R_1\comp\cdots\comp R_m)$-successors which are instances of $C$ in $\A$.
On the other hand, $\I$ satisfies $(\atmostq{n}{R_1\comp\cdots\comp R_m}{C})(x)$
and, thus, there are two different $(R_1\comp\cdots\comp R_m)$-successors
$y_i, y_j$ of $x$ and instances of $C$ in \A\ such that $y_i^\I=y_j^\I$.
Obviously, this implies that $\set{y_i\neq y_j}\cap \A=\emptyset$ and, thus,
$\A'=\A[y_i/y_j]$ is the ABox obtained by applying the $\leq$-rule
to $(\atmostq{n}{R_1\comp\cdots\comp R_m}{C})(x)$. In addition, since
$y_i^\I=y_j^\I$, $\I$ satisfies $\A'$.
Conversely, assume that $\A'=\A[y_i/y_j]$ is obtained from $\A$ by applying the
$\leq$-rule, and let $\I$ be a model of $\A'$. If we consider an interpretation $\I$
so that
$y_j^\I=y_i^\I$ for the additional variable $y_j$ that is present in \A\,
then obviously $\I$ satisfies $\A$.
\item [5. \choos-rule.] Assume that the rule is applied to the constraint
$(\atbothq{n}{R_1\comp\cdots\comp R_m}{C})(x)$ and that
its application yields:
\begin{eqnarray*}
 \A'& = & \A\cup\set{ D(y) }
\end{eqnarray*}
where $D(y)\not\in\A$.
Since \A\ is a subset of $\A'$, any model of  $\A'$ is also a model of \A.
Conversely, assume that $\I$ is a model of \A. As far as
$y$ is concerned,
either $y^\I\in C^\I$ or  $y^\I\in \dom\setminus C^\I=(\sim C)^\I$.
If $y^\I\in C^\I$, for the ABox $\A'$ built with the choice $D=C$
we have that $\I$ satisfies $\A'$. Else, if  $y^\I\in(\sim C)^\I$,
$\I$ satisfies $\A'$ for the choice $D=\sim C$. In any case, $\I$
is a model of the ABox $\A'$ obtained by applying the \choos-rule to \A.
\end{description}
\QED

\noindent\textbf{Proof of Part 2  of Lemma \ref{lemmaX}}
Let \A\ be a complete and clash-free ABox that is obtained by applying the completion rules to  $\set{C_0(x_0)}$.
We define the \emph{canonical interpretation} $\I_\A$ of \A\ as follows:
\begin{enumerate}
  \item The domain $\Delta^{\I_\A}$ of $\I_\A$ consists of all the individual names $x\in\NI$ occurring in \A.
  \item For all concept names $C\in\NC$ we define $C^{\I_\A} := \set{ x\mid C(x)\in\A }$.
  \item For all role names $R\in\NR$ we define $R^{\I_\A} := \set{ (x,y)\mid R(x,y)\in\A }$.
  \item For all individual names $x^{\I_\A}:=x$ (i.e. the variable assignment $\pi$ is the identity on \NI).
\end{enumerate}
We show that $\I_\A$ satisfies every constraint in \A.

By definition, $\I_\A$ satisfies all the role assertions of the form $R(x,y)$,
iff $R(x,y)\in\A$.
More generally, $y$ is an $(R_1\comp\cdots\comp R_m)$-successor of $x$ in \A\ iff
$y$ is an $(R_1\comp\cdots\comp R_m)$-successor of $x$ in $\I_\A$.
Furthermore, $y\neq z$ implies $y^{\I_\A}\neq z^{\I_\A}$ by construction of $\I_\A$.
By induction on the structure
of concept descriptions, it can be easily shown that $\I_\A$ satisfies the concept assertions as well,
provided that \A\ is complete and clash-free. Again, we restrict our attention to number restrictions
and the \choos-rule,
since the induction base and the treatment of other constructors is the same as for \ALC.
\begin{itemize}
  \item First, consider any assertion $(\atbothq{n}{R_1\comp\cdots\comp R_m}{C})(x)\in\A$
and all $y$'s which are
$(R_1\comp\cdots\comp R_m)$-successors of $x$ in \A.
Then, for each of them, either $C(y)\in\A$ or $\sim C(y)\in\A$,
otherwise the  \choos-rule could be applied.
Moreover, it can be easily proved (by induction on the structure of $C$) that
$\set{C(y),\sim C(y)}\subseteq\A$ would lead to a clash.
  \item Consider $(\atleastq{n}{R_1\comp\cdots\comp R_m}{C})(x)\in\A$. Since \A\ is complete,
the $\geq$-rule cannot be applied to $(\atleastq{n}{R_1\comp\cdots\comp R_m}{C})(x)$ and, thus,
$x$ has at least $n$ $(R_1\comp\cdots\comp R_m)$-successors which are instances of $C$ in \A,
which are also $(R_1\comp\cdots\comp R_m)$-successors of $x$ and instances of $C$ in $\I_\A$
(by induction, $y\in C^{\I_\A}$ for each $y$ with $C(y)\in\A$).
Hence, $x\in (\atleastq{n}{R_1\comp\cdots\comp R_m}{C})^{\I_\A}$
  \item Constraints with the form $(\atmostq{n}{R_1\comp\cdots\comp R_m}{C})(x)\in\A$ are
satisfied since \A\ is clash-free and complete. In fact, assume that $x$ has more than $n$
$(R_1\comp\cdots\comp R_m)$-successors which are instances of $C$ in $\I_\A$. Then
$x$ has more than $n$
$(R_1\comp\cdots\comp R_m)$-successors which are instances of $C$ also in \A.
If $\A$ contained
inequality constraints $y_i\neq y_j$ for all these successors, then we would have a clash.
Otherwise, the $\leq$-rule could be applied.
\end{itemize}
\QED

\noindent\textbf{Proof of Part 3 of Lemma \ref{lemmaX}}
Assume that $\A$ contains a clash. If $\set{A(x),(\Not A)(x)}\subseteq \A$, then clearly no
interpretation can satisfy both constraints. Thus assume that $(\atmostq{n}{R_1\comp\cdots\comp R_m}{C})(x)\in\A$ and
$x$ has $p>n$ $(R_1\comp\cdots\comp R_m)$-successors $y_1,\ldots,y_p$ with
$\set{C(y_i)\mid 1\leq i\leq p}\cup\set{y_i\neq y_j\mid 1\leq i<j\leq p}\subseteq \A$. Obviously,
this implies that, in any model $\I$ of \A, $\Int{x}$ has $p>n$ $(R_1\comp\cdots\comp R_m)$-successors
which are instances of $C$ in $\I$, which shows that $\I$ cannot satisfy
$(\atmostq{n}{R_1\comp\cdots\comp R_m}{C})(x)$. \QED

\vskip 5mm
\noindent\textbf{Proof of Part 4 of Lemma \ref{lemmaX}}
We must show that the Tableau algorithm that tests satisfiability of \ALCNX{\comp}-concepts
always terminates. In the following, we consider only ABoxes \A\ that are obtained by applying
the completion rules to $\set{C_0(x_0)}$. For a concept $C$, we define its and/or-size $|C|_{\And,\Or}$
as the number of $\And$ and $\Or$ constructors in $C$. The maximal role depth $\depth(C)$ of $C$ is
defined as follows:
\begin{eqnarray*}
  \depth(A) = \depth(\Not A) & := & 0\quad \text{ for }A\in \NC \\
  \depth(C_1\And C_2) = 
  \depth(C_1\Or C_2) & := & \max\set{ \depth(C_1), \depth(C_2) } \\
  \depth(\atbothq{n}{R_1\comp\cdots\comp R_m}{C}) & = & m + \depth(C)
\end{eqnarray*}
Let $C_0$ be an \ALCNX{\comp}-concept in NNF$^{\Join}$, and let \A\ an ABox obtained
by applying the completion rules to  $\set{C_0(x_0)}$. As an easy consequence
of the definition of the completion rules, we can observe the following facts:
\begin{enumerate}
\item Every variable $x\neq x_0$ that occurs in \A\ is an $(R_1\comp\cdots\comp R_m)$-successor
of $x_0$ for some role chain of length $m\geq 1$. In addition, every other role chain that
connects $x_0$ with $x$ has the same length.
\item If $x$ can be reached in \A\ by a role chain of length $m$ from $x_0$, then for each
constraint $C(x)\in\A$, the maximal role depth of $C$ is bounded by the maximal role depth
of $C_0$ minus $m$ (i.e. $\depth(C)\leq\depth(C_0)-m$). Consequently, $m\leq\depth(C_0)$.
\end{enumerate}
Let $m_0$ be the maximal role depth of $C_0$. Because of the first fact, every individual $x$ in a
ABox \A\ (reached from $\set{C_0(x_0)}$ by applying completion rules) has a unique role level
$\level(x)$, which is its distance from the root node $x_0$, i.e. the unique length of the
role chains that connect  $x_0$ with  $x$. Owing to the second fact, the level of each individual
is an integer between 0 and  $m_0$.

In the following, we define a mapping $K$ of ABoxes \A\ to a $3(m_0+1)$-tuple of non-negative
integers such that $\A\rightarrow\A'$ implies $K(\A)\succ K(\A')$, where $\succ$ denotes
the lexicographic ordering on tuples. Since the lexicographic ordering is well-founded,
this implies termination of the algorithm. In fact, if the algorithm did not terminate, then there would
exist an infinite sequence $\A_0\rightarrow\A_1\rightarrow\cdots$, and this would yield an infinite
descending $\succ$-chain of tuples.

Hence, let \A\ be an ABox that can be reached from $\set{C_0(x_0)}$ by applying completion rules.
We define:
\begin{displaymath}
  K(\A) := (\overline\kappa_0,\overline\kappa_1,\ldots,\overline\kappa_{m_0-1},\overline\kappa_{m_0}),
\end{displaymath}
where (sub)tuple $\overline\kappa_\ell=(\kappa_\ell^1,\kappa_\ell^2,\kappa_\ell^3)$ and the components
$\kappa_\ell^i$ are obtained as follows:
\begin{itemize}
  \item $\kappa_\ell^1$ is the number of individual variables $x$ in \A\ with $\level(x)=\ell$.
  \item $\kappa_\ell^2$ is the sum of the and/or sizes $|C|_{\And,\Or}$ of all constraints
$C(x)\in\A$ such that $\level(x)=\ell$ and the $\And$- or $\Or$-rule is applicable to $C(x)$ .
  \item For a constraint $\alpha(x)=(\atleastq{n}{R_1\comp\cdots\comp R_m}{C})(x)\in\A$, let $s$ be
the cardinality of \emph{maximal sets} $\set{y_1,\ldots,y_s }$, such that
$y_i$ is an $(R_1\comp\cdots\comp R_m)$-successor of $x$ ($1\leq i\leq s$),
$\set{C(y_i)|1\leq i\leq s}\subseteq\A$ and $\set{y_i\neq y_j|1\leq i<j\leq s}\subseteq\A$.
Then we associate with the constraint the number $r(\alpha(x))=\max\set{n-s,0}$, representing the number of individuals
that (possibly) still have to be added to \A\ to make the constraint $\alpha(x)$ satisfied, and define $\kappa_\ell^3$
as follows:
\begin{eqnarray*}
\kappa_\ell^3 &=& \sum_{\alpha(x)\in\A, \level(x)=\ell} r(\alpha(x))
\end{eqnarray*}
\end{itemize}
In the following, we show that $\A\rightarrow\A'$ implies $K(\A)\succ K(\A')$ for each
of the completion rules in Fig.~\ref{fig:tabl1}.
\begin{description}
  \item[1. $\And$-rule.] Assume the rule is applied to the constraint $(C_1\And C_2)(x)$,
let $\A'$ be the ABox obtained by its application and let $\ell=\level(x)$.
First we compare $\overline\kappa_\ell$ and $\overline\kappa_\ell'$, i.e. the tuples associated
with level $\ell$ in \A\ and $\A'$, respectively. The first components $\kappa_\ell^1$
and $\kappa_\ell^{1\prime}$ agree since the number of individuals and their levels have not been changed.
For the second component, we have a decrease (i.e. $\kappa_\ell^{2\prime} < \kappa_\ell^2$), since
$|C_1\And C_2|_{\And,\Or}$ is removed from the sum, and replaced by a number that is no larger than
$|C_1|_{\And,\Or}+|C_2|_{\And,\Or}$ (depending on whether the top constructor of $C_1$ and $C_2$
is $\Or$ or $\And$, or another constructor). Since tuples are compared with lexicographic ordering,
a decrease in the second component makes sure that what happens in the third component is irrelevant.
For the same reason, we need not consider tuples $\kappa_q$ for $q>\ell$.
Tuples at levels $q<\ell$ are either unchanged or have their third component decreased
by the application of the rule, since the addition of the new constraints may add
$x$ to one of the maximal sets involved in the $\kappa_q^3$ definition
(e.g. if $(\atleastq{1}{R}{C_1})(x')$ with $R(x',x)\in\A$ but $C_1(x)\not\in\A$,
we might have a decrement in $\kappa_{\ell-1}^3$ when $C_1(x)$ is added to $\A'$).
  \item[2. $\Or$-rule.] This rule can be treated like the $\And$-rule.
  \item[3. $\geq$-rule.] Assume the rule is applied to the constraint
$(\atleastq{n}{R_1\comp\cdots\comp R_m}{C})(x)$,
let $\A'$ be the ABox obtained by its application and let $\ell=\level(x)$.
The first two components of $\overline\kappa_\ell$ remains unchanged.
The third component decreases (i.e. $\kappa_\ell^{3\prime} < \kappa_\ell^3$),
since the new individuals $z_1,\ldots,z_{n-q}$ can now be added to the maximal sets
of explicitly distinct individuals which are instances of $C$ and $(R_1\comp\cdots\comp R_m)$-successors of $x$
used in the computation of $s$.
For this reason, the increase in the first component of tuples of levels larger than $\ell$ is irrelevant
($z_{i2}$'s are added at level $\ell+1$, 
\ldots, $z_{im}$'s at level $\ell+m-1$, and $z_i$'s are added at level $\ell+m$).
Tuples at levels smaller than $\ell$ are either unchanged or have their third component
decreased by the application of the rule.
  \item[4. $\leq$-rule.] Assume the rule is applied to the constraint
$(\atmostq{n}{R_1\comp\cdots\comp R_m}{C})(x)$,
let $\A'$ be the ABox obtained by its application and let $\ell=\level(x)$.
On level $\ell+m$, the first component $\kappa^1_\ell$ decreases, since variable $y_i$ is removed.
Thus, possible increases in other components of $\overline\kappa_\ell$ are irrelevant.
Tuples associated with smaller levels $q<\ell$ remain unchanged or decrease. In fact, the
third component of tuples of smaller level cannot increase since for the individuals $y_i$ and $y_j$
that have been identified there was no inequality $y_i\neq y_j\in\A$. Moreover,
since no constraints are removed and, in particular, $y_j$ in $\A'$
has all its old constraints plus the constraints of $y_i$ in \A,
$y_i$ may contribute to one of the maximal sets involved in the $\kappa_q^3$ definition
(e.g. if 
$\set{(\atleastq{1}{R}{C})(x'),R(x',y_i),C(y_j)}\subseteq\A$
we might have a decrement in $\kappa_{\ell-1}^3$).
  \item[5. \choos-rule.] Assume the rule is applied to the constraint
$(\atbothq{n}{R_1\comp\cdots\comp R_m}{C})(x)$,
let $\A'$ be the ABox obtained by its application and let $\ell=\level(x)$.
Obviously, the first two components remain unchanged at every level.
Tuples at levels $q$ smaller than $\ell+m$ have their third component unchanged
or decreased, since the addition of the constraint $D(y)$ (with $\level(y)=\ell+m$)
may add some new individual 
to some of the maximal sets involved in the $\kappa_q^3$ definition.
\end{description}
\QED

\subsection*{Complexity issues}
The tableau-based satisfiability algorithm proposed above for \ALCQX{\comp}
may require exponential time and space.
The optimization strategies profitably employed for \ALCN and \ALCQ \cite{Tobi01,BS01}
do not seem to be applicable to \ALCNX{\comp} and \ALCQX{\comp}.
As a matter of fact, such strategies rely on the fact that the
underlying Logics have the \emph{tree model} property, and,
for the sake of satisfiability testing, the individuality
of different role-successors of a given domain object is not relevant.
Only the number of such successors counts (for $\geq$- and $\leq$-rule
applicability and clash testing) and, thus, a single successor
at a time can be used as ``representative'' also for its siblings,
when continuing the algorithm for its further role-successors.
In such a way, only one branch of the tree model at a time can be generated
and investigated by the algorithm, giving rise to a non-deterministic
procedure consuming only polynomial space and, thus, to \PSPACE complexity
(since \NPSPACE=\PSPACE, owing to Savitch's Theorem \cite{Savi70}).
In our case, such an optimization does not seem to be possible, since \ALCNX{\comp}
and \ALCQX{\comp} do not have the tree model property, as number restrictions
$\atleast{p}{R_1\comp\cdots\comp R_{m-1}}\And\atmost{q}{R_1\comp\cdots\comp R_{m-1}\comp R_m}$
(with $p>q$) make some separate $(R_1\comp\cdots\comp R_{m-1})$ role chains merge
into confluent $(R_1\comp\cdots\comp R_{m-1}\comp R_m)$ chains to respect both kinds of number restrictions.
In fact, the identifications of successors effected by the
$\leq$-rule (say at level $\ell$) may involve individuals generated by previous
executions of the $\geq$-rule for different  $\atleastq{n}{R_1\comp\cdots\comp R_m}{C}(x)$
constraints, with possibly different values of $\level(x)$
and role chain lengths (with the proviso that $\level(x)+1\leq\ell\leq\level(x)+m$).
The enforcement of mutual constraints between possibly ``intersecting''
role chains strictly relies on the \emph{individuation} of single successors,
and cannot be surrogated, in general, via representatives.
As a result, the algorithm in Fig.~\ref{fig:tabl1} is a non-deterministic procedure
possibly producing complete ABoxes of \emph{exponential size} in the length of the
input concept description (even if binary coding of numbers is assumed).
\begin{lemma}\label{lemmaK}
Given a complete ABox $\A$ generated by the algorithm in Fig.~\ref{fig:tabl1},
the size of $\A$ is exponential in the input size $s$, thanks to the
following facts:
\begin{enumerate}
\item The number $a$ of individuals in $\A$ is $O(2^{p(s)})$, where $p$ is a polynomial
function.
\item The number of constraints in $\A$ is a polynomial function of $a$.
\end{enumerate}
\end{lemma}
Let us define the \emph{size} $c_0$ of the concept description $C_0$ as the total
number of symbols (operators, concept and role names) it contains, and let
$N:=\max\set{n| (\atleastq{n}{R_1\comp\cdots\comp R_m}{C}) \text{ is a subconcept of }C_0}$.
Moreover, the number of subconcepts of $C_0$ is obviously bounded by $c_0$.

\vskip 5mm
\noindent\textbf{Proof of Fact 1}
According to Fig.~\ref{fig:tabl1}, new individuals (apart from $x_0$) are added to $\A$
by the application of the $\geq$-rule only.
The algorithm execution generates a connected structure with the shape of a tree, rooted on $x_0$,
where some node coincide (owing to $\leq$-rule applications).
Each path in this tree-structure has a maximal length which is bounded
by the maximal role depth $m_0$ of $C_0$. The out-degree is bounded by the maximal
number of new successors that can be generated from an individual $x$.
This number cannot exceed $c_0 N m_0$, since the number of
of times the $\geq$-rule can be applied to a constraint on $x$ is limited by
the total number of $\exists^{\geq n}$ constructors in $C_0$ and, thus, by $c_0$
and, for each application of the $\geq$-rule, no more than $N m_0$
new individuals can be added. Hence, the total size of the tree-structure
is bounded by $(c_0  N  m_0)^{m_0}$ $= 2^{m_0(\log c_0+\log N+\log m_0)}$
$\leq 2^{2 c_0^2+c_0\log N}$, since $m_0\leq c_0$. Obviously,
the exponent is a polynomial function of the input size, even if binary
coding of numbers is adopted.
\QED

\vskip 5mm
\noindent\textbf{Proof of Fact 2}
For each individual $x$, $\A$ may at most contain a pair of constraints
$\set{C(x),\sim C(x)}$ for each subconcept $C$ of $C_0$. Hence, the total
number of constraints with the form $D(x)$ in $\A$ is bounded by $2 c_0$.
Moreover, for each pair of individuals $x$ and $y$,
the number of constraints with the form $R(x,y)$ (or $x\neq y$) in $\A$
is limited by the number of role names in $C_0$, which is strictly less than $c_0$,
plus one (for inequality constraints).
Hence, the size of $\A$ is surely bounded by $2 c_0 a + c_0 a^2$
(we could derive a tighter bound if we took into account the role levels of individuals).
\QED

\vskip 5mm
As it can be easily seen, the two facts together give a space consumption
bounded by $2^{6s^2+s}$.

\begin{corollary}
By the given algorithm, deciding satisfiability (subsumption) of \ALCQX{\comp} concepts
is in the \NEXPTIME (\emph{co-}\NEXPTIME) complexity class.
\end{corollary}

\subsection{An extension of the decidability result}
We provide in this Section an extension of the algorithm given in Fig.~\ref{fig:tabl1}
for \ALCQX{\comp}-concept satisfiability,
such that it can also deal with union and/or intersection of role chains \emph{of the same length}.
The extension follows the same directions of the similar extension proposed
for \ALCNX{\comp} in \cite{BS99}. Analogously, also the soundness, completeness and
termination proofs of our extended algorithm are very similar to the ones proposed
for the basic algorithm in the previous Section and, thus, they will only be sketched.

The general form of a role expression $\cal R$ we consider here is the following:
\begin{eqnarray*}
  {\cal R} &=&
\overset{M}{\bigOr{i=1}} \overset{N_i}{\bigAnd{j=1}} (R^{ij}_1\comp R^{ij}_2\comp\cdots R^{ij}_m)
\end{eqnarray*}
that is we assume, for the sake of simplicity, Boolean role chain combinations
to be in Disjunctive Normal Form\footnote{
General $\And/\Or$ combinations of role chains can be put in DNF (which may require an exponential time)
by rewriting concept $C_0$ before the execution of the satisfiability algorithm.} (DNF).
In the presence of role expressions of this kind, we modify the definition
of role successor for a complex role chain $\cal R$ as follows.

\begin{figure}
\hrule
\begin{eqnarray*}
\geq'\text{-rule: } & \text{\bf if}\quad 1. & (\atleastq{n}{{\cal R}}{C})(x)\in\A \text{ and} \\
                   &        \;\;\;\quad 2. & x \text{ has exactly $p$ $\cal R$-successors $y_1,\ldots,y_p$ with }p<n\\
                                    && \text{such that } \set{C(y_i)\mid 1\leq i\leq p}\cup\set{y_i\neq y_j\mid 1\leq i<j\leq p}\subseteq \A\\
                   & \text{\bf then} & \A':=\A\cup\set{ R^{\hat\imath_i j}_1(x,z^j_{i2}),R^{\hat\imath_i j}_2(z^j_{i2},z^j_{i3}),\ldots,R^{\hat\imath_i j}_m(z^j_{im},z^j_i),C(z_i) \\
&&\qquad\qquad\qquad\; \mid 1\leq i\leq n-p, 1\leq j\leq N_{\hat\imath_i} }\\
                                 &&\qquad\quad\;\;    \cup\set{ z^\ell_i\neq z^\ell_j\mid 1\leq i<j\leq n-p, 1\leq\ell\leq N_{\hat\imath_i}} \\
                                 &&\qquad\quad\;\;    \cup\set{ y_i\neq z^\ell_j\mid 1\leq i\leq p, 1\leq j\leq n-p, 1\leq\ell\leq N_{\hat\imath_i}}\\
                                 && \text{for some $\set{\hat\imath_1,\hat\imath_2,\ldots,\hat\imath_{n-p}} \subseteq\set{1,\ldots, M}$,}\\
                                 &&\text{where $z^j_{ik},z^j_i$ (for $1\leq i\leq n-p,2\leq k\leq m,1\leq j\leq N_{\hat\imath_i}$)}\\
                                 &&\text{are $\textstyle m\sum_{i=1}^{n-p}N_{\hat\imath_i}$ fresh variables } \\
\leq'\text{-rule: } & \text{\bf if}\quad 1. & (\atmostq{n}{{\cal R}}{C})(x)\in\A \text{ and} \\
                   &        \;\;\;\quad 2. & x \text{ has more than $n$ $\cal R$-successors $y_1,\ldots,y_p$ such that}\\
 && \set{ C(y_i)\mid 1\leq i\leq p }\subseteq \A \text{ and } \set{y_i\neq y_j}\cap\A=\emptyset \text{ for some $i,j$ ($1\leq i<j\leq p$), } \\
                   & \text{\bf then} & \text{for some pair }y_i,y_j \,(1\leq i<j\leq p) \text{ such that } \set{y_i\neq y_j}\cap\A=\emptyset \\
&& \A':=[y_i/y_j]\A \\
\text{\choos$'$-rule: } & \text{\bf if}\quad 1. & (\atbothq{n}{{\cal R}}{C})(x) \in\A \text{ and }\\
                   &        \;\;\;\quad 2. &  y \text{ is an $\cal R$-successor of $x$ such that }
                  \set{ C(y), \sim C(y) }\cap \A=\emptyset \\
                   & \text{\bf then} & \A':=\A\cup\set{ D(y) } \text{ for some } D\in\set{ C,\sim C }\\
\end{eqnarray*}
\hrule
\caption{\label{fig:tabl2}
The Completion Rules for \ALCQX{\comp} extended with complex role chains}
\vspace*{3mm}\hrule
\end{figure}
\begin{definition}
The individual $y$ is a $\cal R$-\emph{successor} of $x$ in \A\ (where $\cal R$ is defined as above)
iff for some $\hat\imath$ ($1\leq \hat\imath \leq M$), $\exists y^1_2 y^1_3 \ldots y^1_m y^2_2 y^2_3 \ldots y^2_m$
$y^{N_{\hat\imath}}_2 y^{N_{\hat\imath}}_3 \ldots y^{N_{\hat\imath}}_m$ variables in \A\ such that
$\set{ R^{\hat\imath j}_k(y^j_k,y^j_{k+1}) \mid 2\leq k\leq m-1, 1\leq j\leq N_{\hat\imath} }\cup
\set{ R^{\hat\imath j}_1(x,y^j_2),R^{\hat\imath j}_m(y^j_m,y) \mid 1\leq j\leq N_{\hat\imath} } \subseteq\A$. 
\end{definition}
Notice that, owing to this definition, role successors in \A\ are also successors in every model $\I$
of \A: if $\I$ satisfies \A, and $y$ is an $\cal R$-successor of $x$ in \A, then $y^\I$
is an  $\cal R$-successor of $x^\I$ in $\I$.

The Tableau algorithm is extended by replacing the completion rules dealing with number restrictions
and the \choos-rule  with the rules shown in Fig.~\ref{fig:tabl2}, so that the new complex role chains
can be managed.

In order to prove that the new algorithm decides concept satisfiability for this \ALCQX{\comp}
extension, we must prove that all four parts of Lemma~\ref{lemmaX} still hold.
\begin{enumerate}
\item \emph{Local correctness} of the $\geq'$-, $\leq'$- and \choos$'$-rules can be shown
as in the proof of Part 1 of the Lemma~\ref{lemmaX}.
\item The \emph{canonical model} induced by a complete and clash-free ABox is defined
as in the proof of Part 2 of the Lemma~\ref{lemmaX}. The proof that this canonical model
satisfies the ABox is also similar to the one provided for Lemma~\ref{lemmaX}.
Note that the definition we used for $\cal R$-successors coincides with the notion
of  $\cal R$-successors in the canonical model $\I_\A$ induced by \A.
\item The proof that an ABox containing a clash is unsatisfiable is the same as the one
given above. This follows from the fact that role successors in an ABox \A\ are also
successors in every model $\I$ of \A.
\item The proof of \emph{termination} is also very similar to the one considered before.
The definition of the depth of a concept is extended in the obvious way to expressions
involving complex roles:
\begin{eqnarray*}
  \depth(\atbothq{n}{{\cal R}}{C}) =
  \depth\bigg(\atbothq{n}{ \big( \overset{M}{\underset{i=1}\Or} \overset{N_i}{\underset{j=1}\And}
  (R^{ij}_1\comp R^{ij}_2\comp\cdots R^{ij}_m)\big) }{C}\bigg) & = & m + \depth(C)
\end{eqnarray*}
Since role chains in complex roles are all of the same length, the two facts stated in the proof
of Part 4 of Lemma~\ref{lemmaX} are still valid and, thus, we can define the same metric $K(\A)$
as before also on all the ABoxes that are produced by the new completion rules. It can be seen
that the proof that $\A\rightarrow\A'$ implies $K(\A)\succ K(\A')$ carries over to the new rules.
Actually, the proof given in Part 4 of Lemma~\ref{lemmaX} only relies on the fact that all role
chains connecting any two individuals have the same length, which is still satisfied in the
extended logics.
\end{enumerate}
An immediate consequence of these observations is the Theorem that follows:
\begin{theorem}
Concept satisfiability (and subsumption) for the logic that extends \ALCQX{\comp}
with union/intersections of role chains of the same length is decidable,
and the Tableau algorithm based on the completion rules in Fig.~\ref{fig:tabl2}
is an effective decision procedure.
\end{theorem}

As far as complexity of the algorithm is concerned, Lemma~\ref{lemmaK} holds
also for the algorithm in Fig.~\ref{fig:tabl2}. The only modification required
is to the proof of Fact 2, in the tree-structure out-degree evaluation, since
the application of each $\geq'$-rule may generate at most $N m_0 \hat{N}$
successors, where $\hat N$ is the maximal number of conjuncts occurring
in a role chain combination. Since $\hat N\leq c_0$, the number of individuals
in $\A$ is now bounded by $2^{3 c_0^2+c_0\log N}$.
\begin{corollary}
By the given algorithm, deciding concept satisfiability (subsumption)
for the logic that extends \ALCQX{\comp} with union/intersections of role chains of the same length
is in the \NEXPTIME (\emph{co-}\NEXPTIME) complexity class.
\end{corollary}

\newcommand{\blu}[1]{\textcolor{blue}{#1}}
\newcommand{\rosso}[1]{\textcolor{red}{#1}}
\newcommand{\giallo}[1]{\textcolor{yellow}{#1}}
\newcommand{\verde}[1]{\textcolor{green}{#1}}
\newcommand{\viola}[1]{\textcolor{magenta}{#1}}

\begin{figure}
\setlength{\unitlength}{.3mm}
\centering
\begin{picture}(500,710)(0,-50)

\xput{250,660}{\rosso{\ALCQX{\comp,^-,\Or,\And}}}
\xput{250,640}{\rosso{\ALCNNX{\comp,^-,\Or,\And}}}
\xput{250,620}{\rosso{\ALCNX{\comp,^-,\Or,\And}}}


\drawline(250,610)(100,530)
\drawline(250,610)(200,530)
\drawline(250,610)(300,530)
\drawline(250,610)(400,530)


\xput{100,520}{\blu{\ALCQX{^-,\Or,\And}}}
\xput{100,500}{\blu{\ALCNNX{^-,\Or,\And}}}
\xput{100,480}{\blu{\ALCNX{^-,\Or,\And}}}

\xput{200,520}{\rosso{\ALCQX{\comp,^-,\Or}}}
\xput{200,500}{\rosso{\ALCNNX{\comp,^-,\Or}}}
\xput{200,480}{\rosso{\ALCNX{\comp,^-,\Or}}}

\xput{300,520}{\rosso{\ALCQX{\comp,\Or,\And}}}
\xput{300,500}{\rosso{\ALCNNX{\comp,\Or,\And}}}
\xput{300,480}{\rosso{\ALCNX{\comp,\Or,\And}}}

\xput{400,520}{\rosso{\ALCQX{\comp,^-,\And}}}
\xput{400,500}{\rosso{\ALCNNX{\comp,^-,\And}}}
\xput{400,480}{\rosso{\ALCNX{\comp,^-,\And}}}


\drawline(100,470)(0,330)
\drawline(100,470)(100,330)
\drawline(100,470)(200,330)

\drawline(200,470)(100,330)
\drawline(200,470)(300,330)
\drawline(200,470)(500,330)

\drawline(300,470)(200,330)
\drawline(300,470)(300,330)
\drawline(300,470)(400,330)

\drawline(400,470)(100,330)
\drawline(400,470)(400,330)
\drawline(400,470)(500,330)


\xput{0,320}{\blu{\ALCQX{^-,\And}}}
\xput{0,300}{\blu{\ALCNNX{^-,\And}}}
\xput{0,280}{\blu{\ALCNX{^-,\And}}}

\xput{100,320}{\blu{\ALCQX{^-,\Or}}}
\xput{100,300}{\blu{\ALCNNX{^-,\Or}}}
\xput{100,280}{\blu{\ALCNX{^-,\Or}}}

\xput{200,320}{\blu{\ALCQX{\Or,\And}}}
\xput{200,300}{\blu{\ALCNNX{\Or,\And}}}
\xput{200,280}{\blu{\ALCNX{\Or,\And}}}

\xput{300,320}{\ALCQX{\comp,\Or}}
\xput{300,300}{\blu{\ALCNNX{\comp,\Or}}}
\xput{300,280}{\blu{\ALCNX{\comp,\Or}}}

\xput{400,320}{\rosso{\ALCQX{\comp,\And}}}
\xput{400,300}{\rosso{\ALCNNX{\comp,\And}}}
\xput{400,280}{\rosso{\ALCNX{\comp,\And}}}

\xput{500,320}{\rosso{\ALCQX{^-,\comp}}}
\xput{500,300}{\rosso{\ALCNNX{^-,\comp}}}
\xput{500,280}{\ALCNX{^-,\comp}}


\drawline(100,130)(0,270)
\drawline(100,130)(100,270)
\drawline(100,130)(500,270)

\drawline(200,130)(100,270)
\drawline(200,130)(200,270)
\drawline(200,130)(300,270)

\drawline(300,130)(0,270)
\drawline(300,130)(200,270)
\drawline(300,130)(400,270)

\drawline(400,130)(300,270)
\drawline(400,130)(400,270)
\drawline(400,130)(500,270)


\xput{100,120}{\blu{\ALCQX{^-}}}
\xput{100,100}{\blu{\ALCNNX{^-}}}
\xput{100,80}{\blu{\ALCNX{^-}}}

\xput{200,120}{\blu{\ALCQX{\Or}}}
\xput{200,100}{\blu{\ALCNNX{\Or}}}
\xput{200,80}{\blu{\ALCNX{\Or}}}

\xput{300,120}{\blu{\ALCQX{\And}}}
\xput{300,100}{\blu{\ALCNNX{\And}}}
\xput{300,80}{\blu{\ALCNX{\And}}}

\xput{400,120}{\blu{\ALCQX{\comp}}}
\xput{400,100}{\blu{\ALCNNX{\comp}}}
\xput{400,80}{\blu{\ALCNX{\comp}}}


\drawline(250,-10)(100,70)
\drawline(250,-10)(200,70)
\drawline(250,-10)(300,70)
\drawline(250,-10)(400,70)


\xput{250,-20}{\blu{\ALCQ}}
\xput{250,-40}{\blu{\ALCN}}



\xput{20,610}{\rosso{\textsl{UNDECIDABLE}}}
\xput{20,590}{\blu{\textsl{DECIDABLE}}}
\dashline{3}(-50,600)(150,600)(150,460)
(255,310)(350,310)
\dashline{3}(235,340)(350,340)(350,260)(550,260)
\dashline{3}(450,260)(450,290)
\Thicklines
\verde{\path(450,290)(550,290)(550,340)(450,340)(450,290)}
\verde{\path(350,90)(450,90)(450,140)(350,140)(350,90)}

\end{picture}
\caption{\label{fig:status}
Decidability status of concept satisfiability in \ALCN extensions.
The inclusion edges are meant between corresponding DLs
in the $\cal N/\bar N/Q$ stack.
The green boxes highlight the DLs studied in this paper.}
\end{figure}
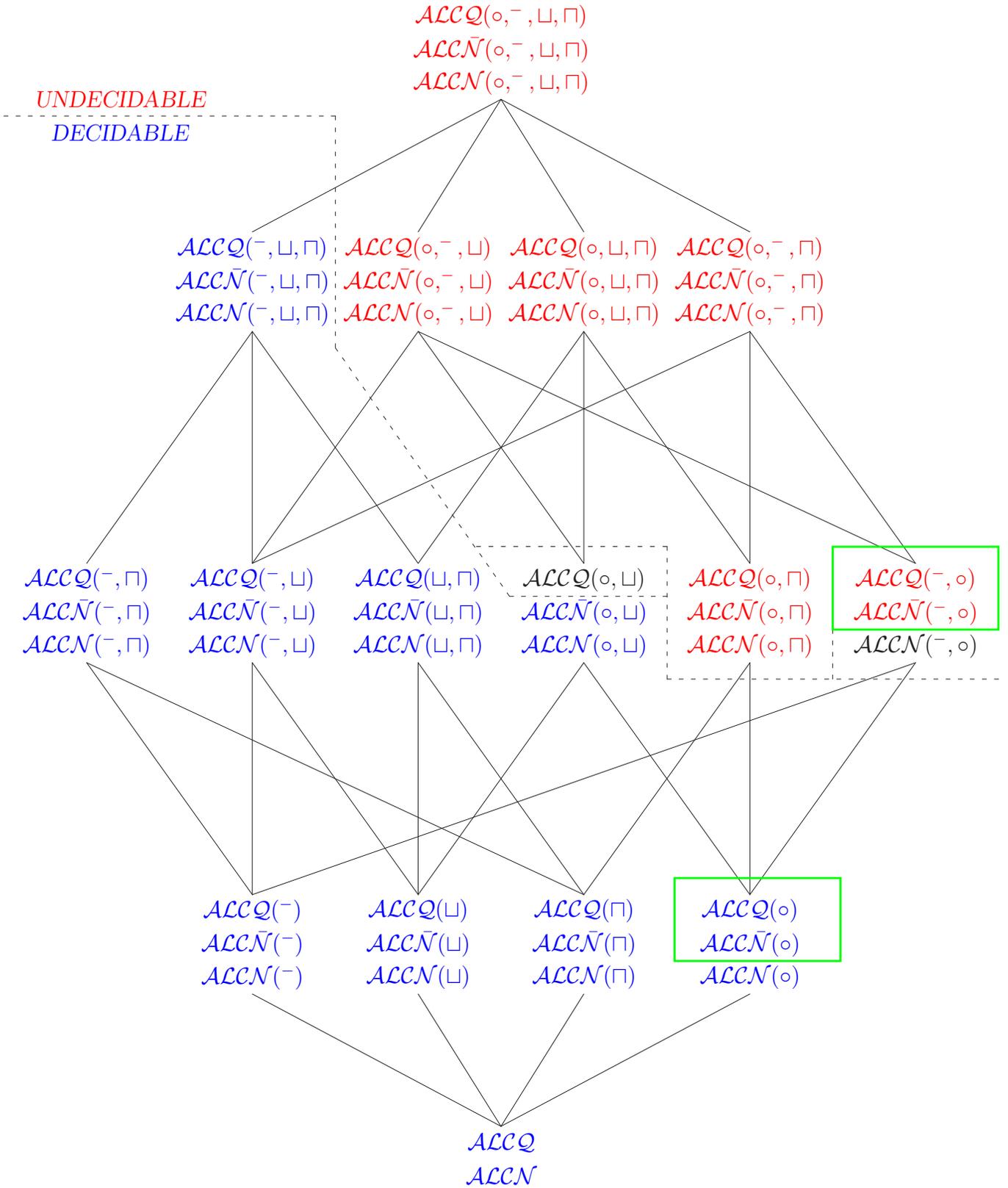

\section{Conclusions}

In this paper we studied expressive Description Logics 
allowing for number restrictions on complex roles built with
the composition operator ($\comp$), 
extended with other role constructors in $\set{\inv{},\Or,\And}$ and qualified number restrictions.

In this framework, we improved the (un)decidability results
by Baader and Sattler on logics of the \ALCN family \cite{BS99}
by showing that \ALCNNX{\comp,^-} is undecidable via reduction of a domino problem,
whereas the introduction of qualified number restrictions in \ALCQX{\comp}
(and in its extension with $\And/\Or$ combinations of role chains with the same length)
does not hinder decidability of reasoning.
For \ALCQX{\comp}, a tableau-based satisfiability algorithm
with a \NEXPTIME upper bound has been proposed.

As we observed in the Introduction that known decidability results
also lift up to \ALCQX{^-,\Or,\And}, we shed some new light on the
whole decidability scenario ranging from \ALCN to
\ALCQX{\comp,^-,\Or,\And}, which is depicted in
Fig.~\ref{fig:status}. In this scenario, since we recently proved
that \ALCNX{\comp,\Or} (for which \ALCNNX{\comp,\Or} is a syntactic variant) is decidable \cite{myDL03},
small gaps left open concern decidability of \ALCQX{\comp,\Or}
and of ``pure'' \ALCNX{\comp,^-}. In particular, around the narrow borders of the second gap,
we proved in this work that the language with inverses in value restrictions and
inverses and composition of roles under unqualified number
restrictions is undecidable, whereas the language with inverses
and role composition under value restrictions and inverses under
qualified number restrictions is decidable, as it is a sublanguage
of the DL $\cal CIQ$ \cite{KR-96}. Another open question is the
exact characterization of \ALCQX{\comp} (and \ALCNX{\comp})
complexity, as the \NEXPTIME bound we derived may be far from
being tight. Future work will also consider such issues.


\begin{thebibliography}{10}

\bibitem{AF99}
{A. Artale and E. Franconi}.
\newblock {Temporal ER Modeling with Description Logics}.
\newblock In {\em Proc. Intl' Conf. on Conceptual Modeling (ER'99)}, pages
  81--95, {Paris, France}, November 1999.

\bibitem{BJ92}
{A. Borgida and M. Jarke}.
\newblock {Knowledge Representation and Reasoning in Software Engineering}.
\newblock {\em IEEE Transactions on Software Engineering}, 18(6):449--450,
  1992.

\bibitem{DLbook}
F.~Baader, D.~McGuinness, D.~Nardi, and P.F. Patel-Schneider, editors.
\newblock {\em The Decsription Logic Handbook: Theory, implementation and
  applications}.
\newblock Cambridge University Press, Cambridge, UK, 2003.

\bibitem{BS99}
F.~Baader and U.~Sattler.
\newblock {Expressive Number Restrictions in Description Logics}.
\newblock {\em J. of Logic and Computation}, 9(3):319--350, 1999.

\bibitem{BS01}
F.~Baader and U.~Sattler.
\newblock {An Overview of Tableau Algorithms for Description Logics}.
\newblock {\em Studia Logica}, 69:5--40, 2001.

\bibitem{domino}
R.~Berger.
\newblock {The Undecidability of the Dominoe Problems}.
\newblock {\em Mem. Amer. Mathematical Society}, 66:1--72, 1966.

\bibitem{C2}
A.~Borgida.
\newblock {On the Relative Expressiveness of Description Logics and First Order
  Logics}.
\newblock {\em Artificial Intelligence}, 82:353--367, 1996.

\bibitem{GHB96}
{C.A. Goble and C. Haul and S. Bechhofer}.
\newblock {Describing and Classifying Multimedia Using the Description Logic
  GRAIL}.
\newblock In {\em Proc. of Storage and Retrieval for Image and Video Databases
  (SPIE IV)}, pages 132--143, {San Diego/La Jolla, CA}, {January/February}
  1996.

\bibitem{dl-expr-reas:book-99}
D.~Calvanese, G.~{De Giacomo}, M.~Lenzerini, and D.~Nardi.
\newblock {Reasoning in Expressive Description Logics}.
\newblock In {\em Handbook of Automated Reasoning}, pages 1581--1634. Elsevier
  Science, {Amsterdam, The Netherlands}, 2001.

\bibitem{JLC-DL-99}
D.~Calvanese, G.~De Giacomo, and M.~Lenzerini.
\newblock {Representing and Reasoning on XML Documents: A Description Logic
  Approach}.
\newblock {\em J. of Logic and Computation}, 9(3):295--318, 1999.

\bibitem{KR-98}
D.~Calvanese, G.~De Giacomo, M.~Lenzerini, D.~Nardi, and R.~Rosati.
\newblock {Description Logic Framework for Information Integration}.
\newblock In {\em Proc. of Intl' Conf. on the Principles of Knowledge
  Representation and Reasoning (KR'98)}, pages 2--13, {Trento, Italy}, June
  1998.

\bibitem{DBIS98}
D.~Calvanese, M.~Lenzerini, and D.~Nardi.
\newblock {Description Logics for Conceptual Data Modeling}.
\newblock In {\em Logics for Databases and Information Systems}, pages
  229--263. Kluwer Academic Publishers, {Boston, MA}, 1998.

\bibitem{ALCN1}
F.M. Donini, M.~Lenzerini, D.~Nardi, and W.~Nutt.
\newblock {The Complexity of Concept Languages}.
\newblock {\em Information and Computation}, 134:1--58, 1997.

\bibitem{myDL01}
{F. Grandi}.
\newblock {On Expressive Number Restrictions in Description Logics}.
\newblock In {\em Proc. of Intl' Workshop on Description Logics (DL'01)}, pages
  56--65, {Stanford, CA}, August 2001.

\bibitem{myDL03}
{F. Grandi}.
\newblock {A Tableau Algorithm for \ALCNX{\comp,\Or}}.
\newblock In {\em Proc. of Intl' Workshop on Description Logics (DL'03)},
  {Rome, Italy}, September 2003.
\newblock \emph{To appear}.

\bibitem{franconi-dood00}
E.~Franconi, F.~Grandi, and F.~Mandreoli.
\newblock {A Semantic Approach for Schema Evolution and Versioning in
  Object-Oriented Databases}.
\newblock In {\em Proc. Intl' Conf. on Deductive and Object-Oriented Databases
  (DOOD 2000)}, pages 1048--1062, {London, UK}, 2000.

\bibitem{KR-96}
G.~De Giacomo and M.~Lenzerini.
\newblock {TBox and ABox Reasoning in Expressive Description Logics}.
\newblock In {\em Proc. of Intl' Conf. on the Principles of Knowledge
  Representation and Reasoning (KR'96)}, pages 348--353, {Cambridge, MA},
  November 1996.

\bibitem{decC2}
E.~{Gr\"{a}del}, M.~Otto, and E.~Rosen.
\newblock {Two-variable Logic with Counting is Decidable}.
\newblock In {\em Proc. Annual IEEE Symp. on Logic in Computer Science
  (LICS'97)}, pages 306--317, {Warsaw, Poland}, 1997.

\bibitem{HB91}
B.~Hollunder and F.~Baader.
\newblock {Qualifying Number Restrictions in Concept Languages}.
\newblock In {\em Proc. of 2nd International Conference on Principles of
  Knowledge Representation and Reasoning ({KR}'91)}, pages 335--346, Cambridge,
  MA, April 1991.

\bibitem{ALCN2}
B.~Hollunder, W.~Nutt, and M.~Schmidt-Schau{\ss}.
\newblock {Subsumption Algorithms for Concept Description Languages}.
\newblock In {\em Proc. of Europ. Conf. on Artificial Intelligence (ECAI'90)},
  pages 335--346, {Stockolm, Sweden}, 1990.

\bibitem{HS01}
{I. Horrocks and U. Sattler}.
\newblock {Ontology Reasoning in the ${\cal SHOQ}(D)$ Description Logic}.
\newblock In {\em Proc. of Intl' Joint Conf. on Artificial Intelligence
  (IJCAI'01)}, pages 199--204, {Seattle, WA}, 2001.

\bibitem{ALC}
{M. Schmidt-Schau{\ss} and G. Smolka}.
\newblock {Attributive Concept Descriptions with Complements}.
\newblock {\em Artificial Intelligence}, 48(1):1--26, 1991.

\bibitem{complC2}
L.~Pacholski, W.~Szwast, and L.~Tendera.
\newblock {Complexity of Two-variable Logic with Counting}.
\newblock In {\em Proc. Annual IEEE Symp. on Logic in Computer Science
  (LICS'97)}, pages 318--327, {Warsaw, Poland}, 1997.

\bibitem{Savi70}
W.~J. Savitch.
\newblock {Relationship between Nondeterministic and Deterministic Tape
  Complexities}.
\newblock {\em Journal of Computer and System Sciences}, 4:177--192, 1970.

\bibitem{Tobi01}
S.~Tobies.
\newblock {\em {Complexity Results and Practical Algorithms for Logics in
  Knowledge Representation}}.
\newblock PhD thesis, RWTH Aachen, Germany, 2001.

\end{thebibliography}

\end{document}